\documentclass{pasa}%

\usepackage{graphicx}

\title[RRAT \rrat\ with the MWA and Parkes]{The emission and scintillation properties of RRAT \rrat\ at 154\,MHz and 1.4\,GHz}

\author[B.~W.~Meyers et al.]{
B.~W.~Meyers$^{1,2,}$\thanks{E-mail: bradley.meyers@postgrad.curtin.edu.au}, 
S.~E.~Tremblay$^{1}$, 
N.~D.~R.~Bhat$^{1}$,
R.~M.~Shannon$^{3,4}$,
S.~M.~Ord$^{2}$,
C.~Sobey$^{5}$,
M.~Johnston-Hollitt$^{1}$,
M.~Walker$^{6}$,
R.~B.~Wayth$^{1}$
\affil{$^{1}$International Centre for Radio Astronomy Research, Curtin University, Bentley, WA 6102, Australia}
\affil{$^{2}$CSIRO Astronomy and Space Science, PO Box 76, Epping, NSW 1710, Australia}
\affil{$^{3}$Centre for Astrophysics and Supercomputing, Swinburne University of Technology, P.O. Box 218, Hawthorn, VIC 3122, Australia}
\affil{$^{4}$ARC Centre of Excellence for Gravitational-wave Discovery (OzGrav), Australia}
\affil{$^{5}$CSIRO Astronomy and Space Science, PO Box 1130, Bentley, WA 6102, Australia}
\affil{$^{6}$Curtin Institute of Radio Astronomy, GPO Box U1987, Perth, WA 6845, Australia}
}

\jid{PASA}
\doi{10.1017/pas.\the\year.xxx}
\jyear{\the\year}

\usepackage{aas_macros}
\usepackage{hyperref} 
\hypersetup{
colorlinks,
citecolor=blue,
linkcolor=blue,
urlcolor=blue
}

\newcommand{\rrat}{J2325$-$0530} % number of MWA + PKS coincident pulses
\newcommand{\nMWA}{89} % number of MWA pulses above 6-sigma
\newcommand{\nPKS}{70} % number of PKS pulses above 6-sigma
\newcommand{\nMatch}{45} % number of MWA + PKS coincident pulses
\newcommand{\specidx}{-2.2\pm 0.1} % mean single-pulse spectral index
\newcommand{\nud}{\nu_{\rm diss}}
\newcommand{\td}{\tau_{\rm diss}}
\newcommand{\cn}{\overline{C^2_{\rm n}}}
\newcommand{\nudPKS}{102\pm 72} % decorrelation bandwidth at 1.4 GHz
\newcommand{\tdPKS}{3478\pm 2550} % decorrelation time scale at 1.4 GHz
\newcommand{\tdMWA}{34\pm 18} % decorrelation time scale at 154 MHz
\newcommand{\vissNE}{44\pm 36} % scintillation speed from 1.4 GHz parameters, using NE2001 distance
\newcommand{\vissYMW}{64\pm 52} % scintillation speed from 1.4 GHz parameters, using YMW distance

\begin{document}

\begin{frontmatter}
\maketitle
\begin{abstract}
Rotating Radio Transients (RRATs) represent a relatively new class of pulsar, primarily characterised by their sporadic bursting emission of single pulses on time scales of minutes to hours. 
In addition to the difficulty involved in detecting these objects, low-frequency ($<300$\,MHz) observations of RRATs are sparse, which makes understanding their broadband emission properties in the context of the normal pulsar population problematic. 
Here, we present the simultaneous detection of RRAT \rrat\ using the Murchison Widefield Array (154\,MHz) and Parkes radio telescope (1.4\,GHz).
On a single-pulse basis, we produce the first polarimetric profile of this pulsar, measure the spectral index ($\alpha=\specidx$), pulse energy distributions, and present the pulse rates in the context of detections in previous epochs.
We find that the distribution of time between subsequent pulses is consistent with a Poisson process and find no evidence of clustering over the $\sim 1.5$\,hr observations.
Finally, we are able to quantify the scintillation properties of RRAT \rrat\ at 1.4\,GHz, where the single pulses are modulated substantially across the observing bandwidth, and show that this characterisation is feasible even with irregular time sampling as a consequence of the sporadic emission behaviour.
\end{abstract}

\end{frontmatter}

\section{INTRODUCTION}\label{sec:intro}
Rotating Radio Transients (RRATs) are a relatively new population of pulsar that were discovered after re-processing of the Parkes Multibeam Pulsar Survey for single-pulse events \citep{2006NatureMcLaughlin}.
They are characterised by sporadic emission of individual pulses, where a single pulse is detected followed by no detectable emission for many rotations (sometimes minutes to hours).
RRATs are almost certainly Galactic neutron stars with extreme emission variability \citep[e.g.][]{2006NatureMcLaughlin,2009MNRASMcLaughlin,2011MNRASKeane,2016MNRASKeane,2018MNRASBhattacharyya}.
Based on objects with adequate observations, we expect single pulse rates in the range of a few pulses to a few hundred pulses per hour.
RRATs are therefore more easily detected through single-pulse searches as opposed to the standard Fourier domain search or traditional folding techniques.
Even though there are 111 known RRATs\footnote{Retrieved from \url{http://astro.phys.wvu.edu/rratalog}}, the inherent difficulty in their detection has meant that the physics responsible for the sporadic nature of the emission remains unclear.

The pulsar and magnetosphere system geometries are thought to play a vital role in the characteristics of pulsar emission, and can be constrained through polarisation measurements \citep[e.g.][]{1998MNRASGould,1998MNRASManchester,1999ApJSWeisberg,2001ApJEverett,2016ApJMitra,2018MNRAJohnston}.
For RRATs this can pose a challenge given that, in general, the folded profiles are not particularly well defined by virtue of their sporadic emission.
Nevertheless, when the polarisation properties have been analysed, even based on a small sample of single pulses, they provide remarkable insight into the nature of the emission (e.g. RRAT J1819$-$1458, \citealp{2009MNRASKarastergiou}).
Generally speaking, very little is known about whether the RRAT population exhibits polarisation characteristics similar to the normal pulsar population.
This is, in part, due to a lack of single-pulse analysis of normal pulsars in the literature, combined with the difficulty of creating high quality polarimetric profiles of RRATs. 

Several models have been proposed to explain the sporadic emission, most of which are also linked to intermittent pulsars and the nulling phenomenon.
Some examples include: extreme nulling \citep[e.g.][]{2007MNRASWang,2010MNRASBurke-Spolaor}, asteroidal or circumpulsar debris \citep{1981ApJMichel,2006ApJLi,2008ApJCordes}, or even mechanisms within the pulsar magnetosphere \citep[e.g.][]{2010MNRASTimokhin,2012ApJLi,2014MNRASMelrose}.
Studying the pulse-energy distributions \citep[e.g.][]{2018ApJShapiro-Albert,2018MNRASMickaliger}, timing periodicities and pulse clustering \citep[e.g.][]{2011MNRASPalliyaguru}, and flux density or pulse energy correlations with single-pulse detection statistics \citep[e.g.][]{2017ApJCui} of RRATs is vital in understanding their emission and how they connect to the canonical pulsar population.

To further uncover connections between RRATs and normal pulsars it is also important to understand their kinematic properties, such as space velocities and proper motions. 
Techniques used to do this for normal pulsars include Very Long Baseline Interferometry (VLBI) \citep[e.g.][]{2018ApJDeller}, multi-wavelength analysis of binary systems \citep[e.g.][]{2018ApJJennings}, long-term precision timing experiments \citep[e.g.][]{2010A&AJanssen,2011ApJGonzalez}, and scintillation analysis \citep[e.g.][]{1986ApJCordes,1998MNRASJohnston,2018ApJSBhat}.
None of these techniques have been applied to RRATs in order to extract the pulsar velocities, specifically.
In particular cases, scintillation studies of RRATs would nominally be able to not only provide estimates of the space velocities, but also allow direct measurement of the turbulence and characteristic scales of the interstellar medium (ISM) along the respective sight-lines, thus allowing them to be used as additional probes of the structure and composition of the ISM.

RRAT \rrat\ was originally discovered as part of the Robert C. Byrd Green Bank Telescope (GBT) 350\,MHz Drift-scan pulsar survey \citep{2013ApJBoyles,2013ApJLynch}.
The pulsar has a pulse period of $P=0.868$\,s, a moderate dispersion measure, $\rm DM=14.966\pm 0.007\,pc\,cm^{-3}$ and a nominal pulse rate of $\rm\sim 50\,hr^{-1}$.
\citet{2015ApJKarako-Argaman} conducted follow-up observations of a subset of those RRATs detected in the survey, including RRAT \rrat, using the GBT at 350\,MHz (though with a larger bandwidth and upgraded digital backend) and the Low Frequency Array (LOFAR; \citealp{2013A&AvanHaarlem,2011A&AStappers}) core stations at 150\,MHz.
This pulsar has also been observed with the first station of the Long Wavelength Array \citep[LWA1;][]{2012JAITaylor} over a frequency range of 30--80\,MHz \citep{2016ApJTaylor}, allowing the measurement of a relatively shallow spectral index ($\alpha_{30}^{80}\approx -0.7$).

In this paper, we present simultaneous observations of single pulses from RRAT \rrat\ with the Murchison Widefield Array (MWA) at 154\,MHz and Parkes radio telescope at 1.4\,GHz.
RRAT \rrat\ is also the first RRAT detected with the MWA.
In Section~\ref{sec:obs} we describe the observations and calibration procedures.
Section~\ref{sec:analysis} presents our analysis and results, followed by discussion in Section~\ref{sec:discussion}.
Finally, we summarise in Section~\ref{sec:conclusions}.
Throughout, we define the spectral index, $\alpha$, by $S_{\nu}\propto\nu^{\alpha}$, where $S_{\nu}$ is the flux density measured at frequency $\nu$.

\section{OBSERVATIONS AND CALIBRATION}\label{sec:obs}
RRAT \rrat\ was simultaneously observed with the MWA and Parkes radio telescope on 2017 June 27.
The MWA observed with a 30.72\,MHz bandwidth centred on 154.24\,MHz for 1.4\,hours, while Parkes observed at a centre frequency of 1396\,MHz with 256\,MHz bandwidth for 1.6\,hours. 
Observing details are summarised in Table~\ref{tab:obs}.

\begin{table*}
\caption{Observing details for MWA and Parkes on 2017 June 27. \label{tab:obs}}
\centering
\begin{tabular}{@{}lcc@{}}
\hline\hline
Parameter & MWA & Parkes \\
\hline
Centre frequency (MHz) & 154.24 & 1396\\
Bandwidth (MHz) & 30.72 & 256\\
Time resolution (ms) & 0.1 & 0.256\\
Channel bandwidth (MHz) & 0.01 & 0.5\\
UTC start time  & 20:30:05 & 20:30:22\\
Observation duration (s) & 4399 & 5867\\
\hline
Dispersion smearing in lowest channel (ms) & 0.46 & 0.03\\
Dispersion delay across bandwidth (ms) & 1060.58 & 11.88\\
Dispersion delay between observed bands$^a$ (ms) & \multicolumn{2}{c}{3192.48}\\
\hline\hline
\end{tabular}
\tabnote{$^a$Delay between the highest Parkes band and the lowest MWA band.}
\end{table*}

\subsection{MWA}\label{sec:mwa_obs}
The MWA is a low-frequency (70--300\,MHz) Square Kilometre Array (SKA) precursor telescope.
Phase I of the MWA was composed of 128 tiles, each containing 16 dual-polarisation dipole antennas, distributed with a maximum baseline of $\sim 3$\,km \citep{2013PASATingay}.
The Phase II upgrade of the MWA, which was completed in October 2017, provides an additional 128 tiles: 76 in two redundant hexagonal configurations near the array centre, with the remaining 52 tiles facilitating maximum baselines of $\sim 6$\,km \citep{2018PASAWayth}.
Presently, hardware constrains the MWA software correlator \citep{2015PASAOrd} to only ingest dual-polarisation inputs from 128 tiles at a time, thus the MWA Phase II is periodically reconfigured between a \textit{compact} and \textit{extended} layout, as described by \citet{2018PASAWayth}.
Our observations were taken in the compact configuration, which was operationally complete as of October 2016. 
 
The Voltage Capture System (VCS) is the high time and frequency resolution recording system for the MWA \citep{2015PASATremblay}.
It records the polyphase filter bank channelised voltages from both polarisations of every connected tile.
This provides critically sampled tile voltages, with a time resolution of 100\,$\mu$s and frequency resolution of 10\,kHz from each of the $24\times 1.28$\,MHz coarse channels that constitutes the full 30.72\,MHz bandwidth.
These data stream to on-site disks at a rate of $\rm\sim 28\,TB\,hr^{-1}$, where they are then automatically transferred to the MWA data archive hosted at the Pawsey Supercomputing Centre, in Perth, Western Australia.
For this observation, we recorded data at a centre frequency of 154.24\,MHz with a bandwidth of 30.72\,MHz for 5153 seconds.

\subsubsection{Tied-array beamforming}
The VCS data were processed offline at the Pawsey Supercomputing Centre on the Galaxy cluster.
A tied-array (coherent) beam is formed by summing the tile voltages in phase and then converted into power (\citealp{2019arXivOrd}, also see \citealp{2016ApJBhat,2017ApJMeyers,2018ApJSBhat}).
This post-processing operation reduces the field-of-view to approximate the size of the synthesised beam of the array ($\sim$1.4\,arcmin in the extended configuration, and $\sim$28\,arcmin in the compact configuration at 150\,MHz).
The tied-array beamforming process provides a boost in sensitivity compared to the incoherent sum; where tile powers are directly summed, and we preserve the wide field-of-view from the tile beam.
While less sensitive, the incoherent sum is nominally a more robust measurement as it requires far fewer post-processing steps, does not depend on adequate convergence of calibration solutions, and is less affected by ionospheric distortions.
Nevertheless, the theoretical improvement of a coherent beam compared to an incoherent sum is $\sqrt{N}$, where $N$ is the number of elements used to create the coherent beam, and provides the maximum sensitivity achievable with the VCS.

In order to create the tied-array beam, calibration information is produced by the Real Time System \citep[RTS;][]{2008ISTSPMitchell}.
The tied-array beamforming software takes the RTS output solutions (i.e. the tile polarimetric response model, complex amplitudes and gains, on a per tile, per coarse channel basis) and computes the necessary cable and geometric delays to point the tied-array beam at the desired position.
For this observation, calibration solutions were created from an observation of PKS 2356$-$61, approximately 2 hours after the observation of RRAT \rrat.
The output from the tied-array beamforming software is full Stokes search-mode PSRFITS data, with the native VCS time and frequency resolution.

\subsubsection{Flux density calibration}\label{sec:mwa_flux}
To determine the system temperature and gain for the tied-array beam, we followed the procedure developed by \citet{2017ApJMeyers}.
Briefly, this involves simulating the tied-array beam pattern by computing the MWA tile beam \citep{2015RaScSutinjo} and multiplying it by the array factor, which incorporates information about individual tile positions and desired pointing direction (see eq.~11 and eq.~12 of \citealp{2017ApJMeyers}). 

The beam pattern is then multiplied in image-space with the radio-frequency global sky model of \citet{2008MNRAS_GSM} and integrated over the visible sky to determine the antenna temperature, $T_{\rm ant}$.
The total system temperature is given by $T_{\rm sys} = \eta T_{\rm ant} + T_{\rm rec}$, where $T_{\rm rec}=34$\,K is the receiver temperature at 154\,MHz, and $\eta=1$ is the nominal radiation efficiency of the array.
The gain is computed by integrating over the tied-array beam pattern itself to determine the effective collecting area ($A_{\rm e}$), which is then converted into a gain by $G=(A_{\rm e}/2k_{\rm B})\times 10^{-26}$\,K\,Jy$^{-1}$.
For this observation we estimate $T_{\rm sys}=274$\,K and $G=0.33$\,K\,Jy$^{-1}$.

The system equivalent flux density is nominally given by ${\rm SEFD}=T_{\rm sys}/G$.
However, this assumes perfect coherence in the simulation, where the sensitivity increases exactly as $\sqrt{N}$ (a factor of $\sim 11$ when all 128 tiles are combined).
This improvement is generally not achieved due to calibration errors, including an imperfect knowledge of the beam pattern, with typical improvement factors of $\sim 5\text{--}9$ (though see \citealp{2016ApJBhat}).
To correct this, we take the brightest MWA pulse from the tied-array beam data, and compare the signal-to-noise ratio to its counterpart in the incoherent sum (which is not affected by this coherence error) and scale the flux densities accordingly (see eq.~2 of \citealp{2017ApJMeyers}).
Incorporating this correction, the SEFD of the MWA tied-array beam was effectively $\sim 1.7$\,kJy.

\subsection{Parkes}\label{sec:pks_obs}
We observed RRAT \rrat\ with the central beam of the 20-cm multibeam receiver on the 64-m Parkes radio telescope, recording at a centre frequency of 1396\,MHz with 256\,MHz bandwidth.
The observation started on 2017 June 27 20:30:22 UTC and lasted for 5867 seconds.
Data were collected with the Parkes Digital Filter Bank Mark-4 (PDFB4) backend, producing $512\times 0.5$\,MHz frequency channels across the band.
The data were recorded in polarimetric search-mode, where the receiver coherency products were detected and averaged to a time resolution of 256\,$\mu$s and written to disk.

\subsubsection{Flux density and polarisation calibration}\label{sec:pks_flux}
Flux density calibration was achieved by observing the radio galaxy Hydra A (3C 218) as per the normal Parkes Pulsar Timing Array procedure \citep{2013PASAManchester}.
Polarisation calibration was conducted by injecting a linearly polarised signal into the feed, which allows us to measure the differential gain and phase.
We also corrected the cross-coupling and ellipticity of the multibeam feed receptors using a model of the full instrumental response \citep[e.g.][]{2004MNRASOrd}.
These calibration solutions were derived and applied using standard PSRCHIVE tools \citep{2004PASAHotan,2012AR&TvanStraten}.
The nominal SEFD throughout the observation was $\approx 36$\,Jy.

\section{ANALYSIS AND RESULTS}\label{sec:analysis}
\subsection{Single pulse detection}\label{sec:sp}
Both the MWA and Parkes data sets were processed using the DSPSR software package \citep{2010PASAvanStraten}, which subdivided the data into single-pulse time series, with 2048 bins across the pulse period, and were incoherently dedispersed using the catalogued dispersion measure ($\rm 14.966\,pc\,cm^{-3}$).
The data were then processed with the PSRCHIVE routine \textsc{paz} using the median-difference filter to remove the vast majority of radio frequency interference (RFI). 
Additionally, we excised five per cent of each band edge from the Parkes data, and 10 fine channels (each 10\,kHz) for each edge of the MWA 1.28\,MHz coarse channels, where aliasing caused by the polyphase filter bank overlap degrades the data.

To find pulses we used the PSRCHIVE single-pulse finding routine, \textsc{psrspa}, looking for pulses above a signal-to-noise ratio (S/N) threshold of six\footnote{Specifically using the peak finding algorithm {\tt above:threshold=6}}.
This produced a list of 162 candidates for Parkes and 188 candidates for the MWA.
A significant fraction of these candidates were detections within the same pulsar rotation (i.e. peaks above the respective telescope's detection threshold), thus, after filtering for unique pulses, there were 102 detected with Parkes and 89 detected with the MWA.
The time and frequency characteristics of the remaining candidates were visually inspected, which resulted in the removal of a further 32 candidate pulses from the Parkes data.
These final excisions were due to RFI that was not automatically removed in the earlier processing steps.

\begin{figure*}
\centering
\includegraphics[width=\linewidth]{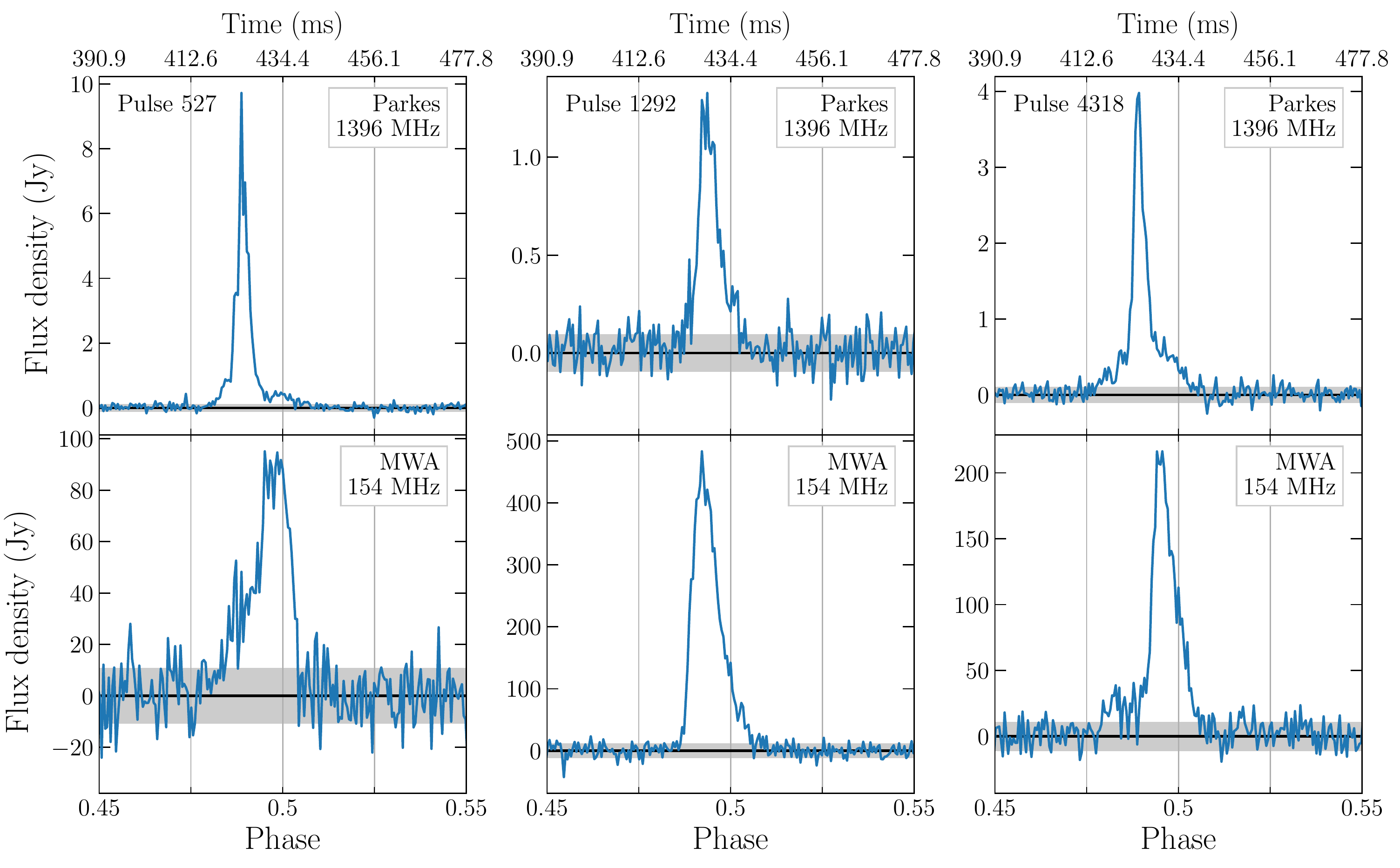}
\caption{Examples of coincident single pulses from RRAT \rrat\ at 1.4\,GHz (Parkes; top row) and 154\,MHz (MWA; bottom row). 
The pulses have been absolutely aligned, in that the same ephemeris was used to reduce the data sets.
The number of rotations since the first simultaneously observed rotation of the pulsar are also given for each pair.
Pulse 527 is the brightest pulse in the Parkes band of the coincident pulses, while pulse 1292 is the brightest in the MWA band.
Pulse 4318 is a relatively average example of a simultaneous pulse.}
\label{fig:spgal}
\end{figure*}

The final catalogue of pulses contained \nMWA\ and \nPKS\ pulses for the MWA and Parkes, respectively.
At this stage, the corresponding flux density scales (see Sections~\ref{sec:mwa_flux} and \ref{sec:pks_flux}) were applied to each single-pulse time series.
For Parkes this was achieved using the standard PSRCHIVE tools and calibration procedures \citep[see e.g.][]{2012AR&TvanStraten}.
Briefly, this required us to construct polarisation and flux density calibration solutions, and then apply these to the individual single pulse archives using the \textsc{pac} tool.
For the MWA VCS data we evaluated the noise baseline on a per-pulse basis and applied the standard radiometer equation, incorporating the simulated system temperature and gain.
Three examples of simultaneously detected pulses are shown in Figure~\ref{fig:spgal}.
The fluence (pulse energy) is estimated by integrating over the pre-determined on-pulse phase window ($\approx 167\text{--}193$ degrees in pulse longitude, or 400--466\,ms) for every detection.

\subsection{Profiles and polarisation}\label{sec:profs_pol}
\begin{figure}
\centering
\includegraphics[width=\columnwidth]{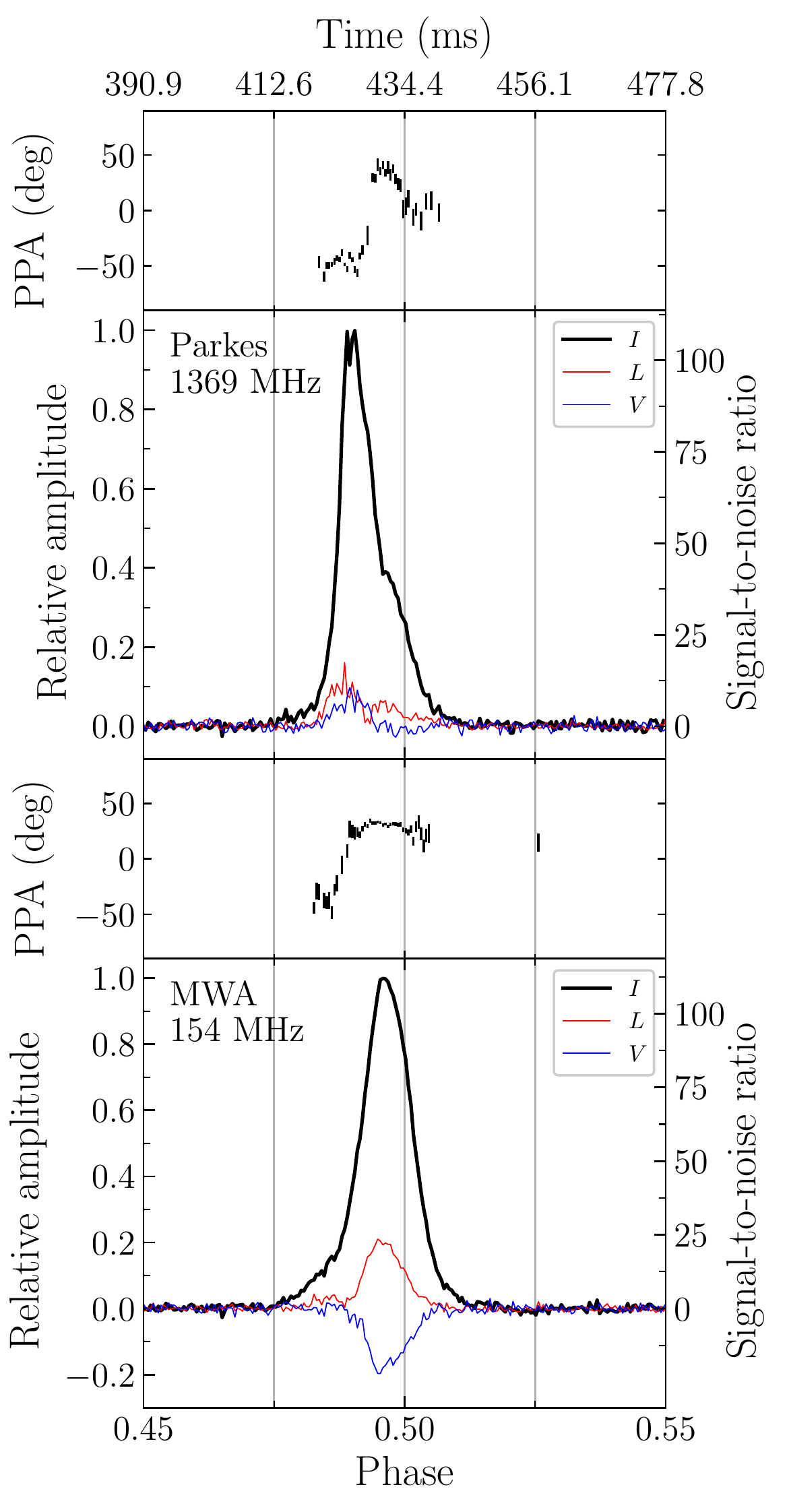}
\caption{A pseudo-integrated profile, combining all single pulses with a $\rm S/N \geq 6$.
The profiles were produced using the same ephemeris and then rotated by 0.5 phase turns.
Total intensity (Stokes $I$) is drawn in black, with linear ($L=\sqrt{Q^2+U^2}$) and circular ($V$) polarisation in red and blue, respectively.
Above each profile is the linear polarisation position angle in degrees.
Both profiles have been corrected for rotation measure (see Section~\ref{sec:rm}).}
\label{fig:profiles}
\end{figure}

We combined the detected pulses into pseudo-integrated profiles which are shown in Figure~\ref{fig:profiles}.
The profiles have been rotated by 0.5 turns for ease of comparison.
No time alignment procedures have been applied to the profiles, thus the profiles are absolutely aligned based on the ephemeris alone.
The ``knee''-like feature in the Parkes profile and the notch at the nominal profile peak are particularly interesting, given that the MWA profile is relativity smooth in comparison\footnote{However, we note that these pseudo-profiles are constructed from less than 100 pulses, whereas pulse profiles typically stabilise only after $\sim 1000$ pulses are averaged \citep[e.g.][]{2012MNRASLiu}.}.
The residual dispersion smearing within the 10\,kHz channels of the MWA data is (at worst) $\sim 0.5$\,ms, which is similar in scale to the Parkes notch feature ($\sim 1$\,ms), thus the smoothness of the MWA profile is possibly an artefact of incoherent dedispersion.
The knee feature in the Parkes profile, and the relatively extended rising edge of the MWA profile are also intriguing.
These profile features would require coherently de-dispersed, high signal-to-noise ratio profiles constructed from many hundreds or thousands of pulses, to examine in detail and to ensure their authenticity. 

The polarisation response of the MWA tied-array beam is currently undergoing self-consistency and cross-validation tests \citep[e.g.][]{2019arXivOrd,2019arXivXue}.
Nonetheless, we present here the first polarisation profile of RRAT \rrat\ at 154\,MHz and 1.4\,GHz.
The profiles have been corrected for Faraday rotation, removing the effects induced by the interstellar medium and ionosphere (see Section~\ref{sec:rm}).
There is clearly substantial polarisation evolution with frequency in this case \citep[see also][]{2019arXivXue}.
Even though the polarisation positional angle (PPA) has not been absolutely calibrated for the MWA, it is reassuring that the general shapes are similar.
For both profiles, we were unable to fit the standard rotating vector model (RVM).
In the case of the MWA profile, one possible reason for this is that scattering induced by the ISM can cause significant deviations from the normally expected RVM (S-like swing) shape \citep[e.g.][]{2009MNRASKarastergiou}.

\subsection{Scintillation}\label{sec:scint}
\begin{figure*}
\centering
\includegraphics[width=\textwidth]{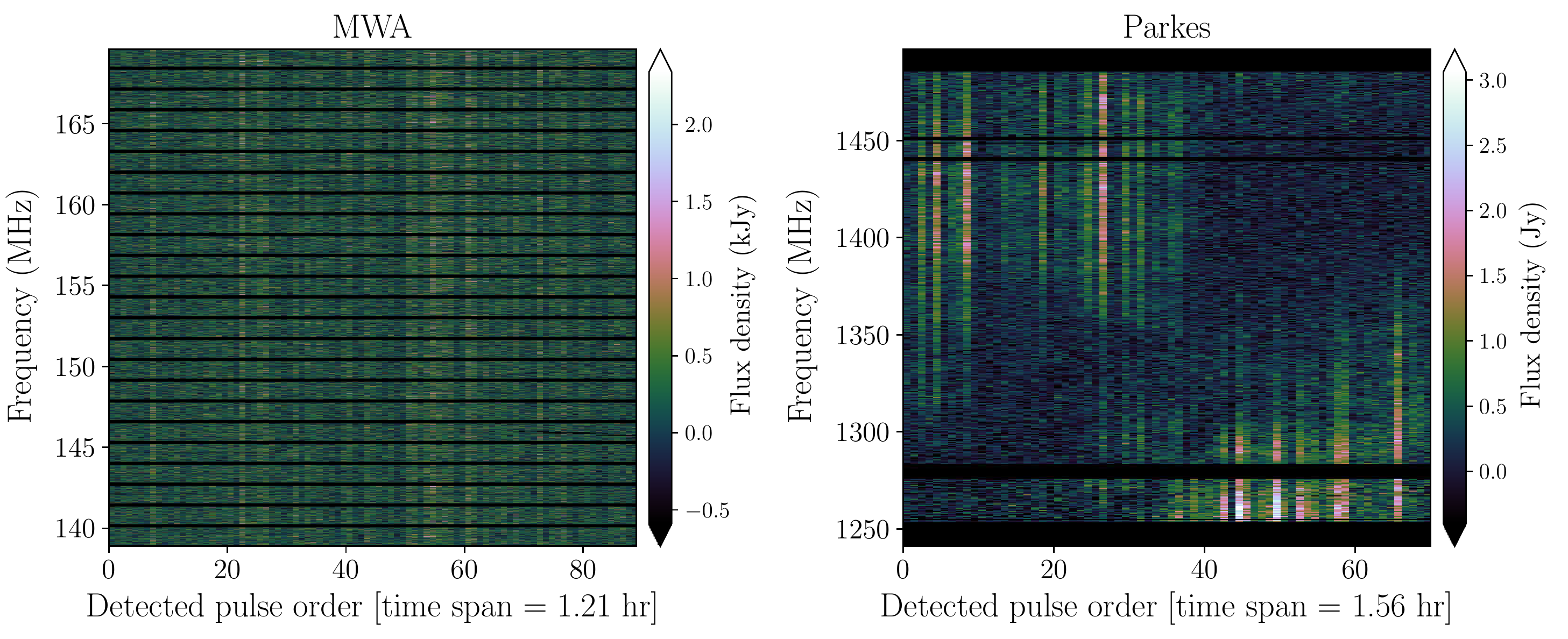}
\caption{A dynamic spectrum of the brightest single pulses from RRAT \rrat\ at 154\,MHz (MWA; left) and 1.4\,GHz (Parkes; right).
The colour scale units are different for each dynamic spectrum (kJy for the MWA data, Jy for the Parkes data), and the $x$-axis represents the order in which the pulses were detected, with the total time spanned by these pulses given for context in the label.
Note that this means the time axis is not continuous (i.e. each column of pixels, corresponding to a single pulse, is not necessarily contiguous with the previous column), unlike standard dynamic spectra.
Nevertheless, it is clear that there is frequency and time structure indicative of diffractive scintillation in the Parkes data, though this is not the case for the MWA data.
The black masked regions are those time and frequency samples excised by the RFI mitigation steps taken during post-processing of the single-pulse data, including coarse channel edges for the MWA, and the colour scale is linear.}
\label{fig:dyn_spec}
\end{figure*}

After combining the single pulses as in Section~\ref{sec:profs_pol}, it was clear that RRAT \rrat\ is affected by diffractive scintillation in the Parkes band.
This was confirmed by examining the dynamic spectrum (see Figure~\ref{fig:dyn_spec}). 
Due to the nature of RRAT emission, the diffractive scintillation pattern is sampled sparsely and irregularly in time, thus performing the standard autocorrelation analysis \citep[e.g.][]{1994MNRASGupta,1999ApJSBhat,2018ApJSBhat} is non-trivial. 
Furthermore, it is difficult to robustly constrain the scintillation parameters given that we only partially sample scintles in time or frequency at 1.4\,GHz (which leads to large statistical uncertainties).
At 154\,MHz it is not immediately clear if there is any scintillation structure present, which suggests that the fine-channel width (10\,kHz) is inadequate to capture the frequency structure. 
The results are summarised in Table~\ref{tab:scint_properties}.
Given these complications, the diffractive scintillation parameters presented here should be considered with caution.

\begin{table*}
\caption{Scintillation properties of RRAT \rrat. \label{tab:scint_properties}}
\centering
\begin{tabular}{@{}cccccc@{}}
\hline\hline
Frequency & $\nud$ & $\td$ & \multicolumn{2}{c}{$V_{\rm iss}$} & $\cn$\\
          &        &       & [$D=0.7$\,kpc] & [$D=1.49$\,kpc]      & \\
(MHz)     & (MHz)  & (s)   & ($\rm km\,s^{-1}$) & ($\rm km\,s^{-1}$) & ($\rm m^{-20/3}$)\\
\hline
154 & $\lesssim 0.01$ & $\tdMWA$ & -- & -- & -- \\
1369 & $\nudPKS$ & $\tdPKS$ & $\vissNE$ & $\vissYMW$ & $\lesssim 2.8\times 10^{-4}$\\
\hline\hline
\end{tabular}
\medskip
\end{table*}

\subsubsection{Scintillation bandwidth}\label{sec:scint_bw}
\begin{figure*}
\centering
\includegraphics[width=\textwidth]{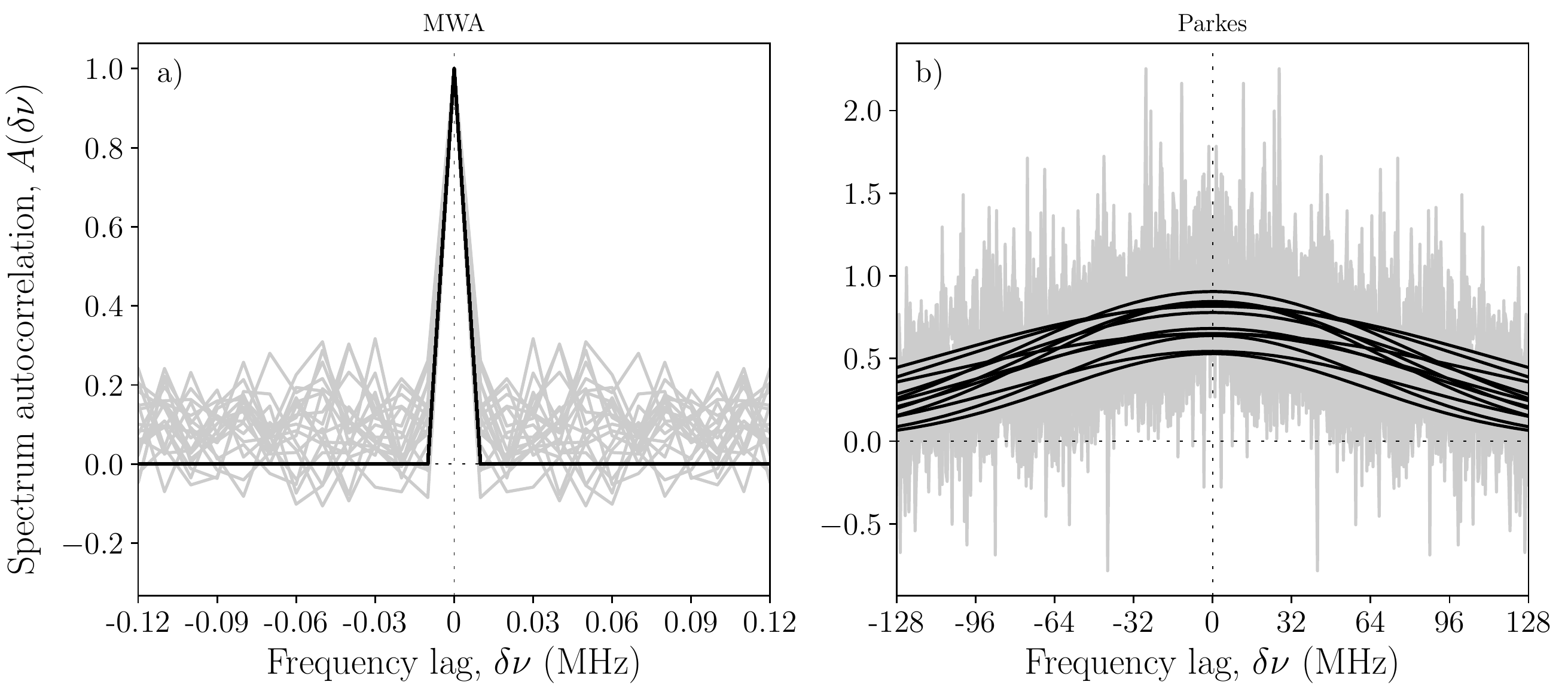}
\caption{The set of ACFs (grey) for the brightest pulses, and their best-fitting Gaussian model (black), from: a) the MWA (15 pulses), and b) Parkes (12 pulses).
The MWA autocorrelations drop to zero by the first frequency lag bin and thus we are not able to even partially resolve the frequency structure. 
From the Parkes data we see structure, though the fact that the autocorrelations do not drop to zero before the last meaningful frequency lag bins indicates that we are not fully sampling a scintle, which is corroborated by the dynamic spectrum in Figure~\ref{fig:dyn_spec}.}
\label{fig:freq_acfs}
\end{figure*}

To estimate the scintillation bandwidth, we measured the mean flux density per frequency channel, $I(t,\nu)$, for every pulse (i.e. the spectrum).
Following \citet{2004ApJCordes}, we then computed the intensity autocorrelation function (ACF),
\begin{equation}
A(\delta\nu) = \langle I(t,\nu)\,I(t,\nu+\delta\nu)\rangle
\end{equation}
for each pulse, where $\delta\nu$ is the frequency lag representing a shift of one channel. 
For each $A(\delta\nu)$ we fit a Gaussian to measure the standard deviation, $\sigma$, and calculate the scintillation bandwidth as $\nud=\left(2\ln2\right)^{1/2}\sigma$ (which corresponds to the half-width at half-maximum of the Gaussian, \citealp[e.g.][]{1986ApJCordes}).
The ACFs and models are normalised by the correlation value corresponding to zero frequency lag, which is calculated as the mean of the correlation value in the adjacent six frequency lag bins (three positive and three negative).
In Figure~\ref{fig:freq_acfs} we show the ACFs and best-fit Gaussian models for the subset of pulses used to estimate the scintillation bandwidth.

Using the above method, we measure an average $\nud=102\pm 12$\,MHz at 1.4\,GHz (i.e. at Parkes) based on the subset of single pulses with ${\rm S/N}>40$ (12/\nPKS\ pulses).
This is nominally a lower limit given that over the observed bandwidth, we do not fully sample even one scintle.
To calculate the expected characteristic scintillation bandwidth at MWA frequencies (154\,MHz), we assume a frequency scaling index of $\gamma=-3.9\pm0.2$ \citep[e.g.][]{2004ApJBhat}, where $\nud\propto\nu^{\gamma}$. 
We find that the expected scintillation bandwidth is $\nud\approx 15$\,kHz.

Processing the single pulses with ${\rm S/N}>20$ (15/\nMWA\ pulses) from the MWA in the same way, we find that in all cases the ACF drops to zero by the first frequency lag bin, indicating that the scintillation bandwidth at 154\,MHz is less than our channel width (i.e. $\nud\lesssim 10$\,kHz).
This indicates that the frequency scaling  is steeper than $\gamma=-3.9$. 
If we take the nominal measured values of $\nud$ at each frequency, and again use the previous defined scaling relation, we find that $\gamma \approx -4.2$, which is steeper than the empirically derived global scaling index \citep{2004ApJBhat}.
We note that nearby pulsars tend to show a steeper scaling index, approaching the extremum (Kolmogorov turbulence) scaling of $\gamma=-4.4$, due to the decreased probability of intervening structures in the ISM that would serve to shallow the scaling index.

Finally, the scintillation characteristics are statistical quantities, thus the sample variance must be included in all quantities derived from the scintillation properties, especially when a small number of scintles are observed. 
This uncertainty is given by $\sigma_{\rm stat} \approx N_{\rm scint}^{-1/2}$, where the number of scintles observed, $N_{\rm scint}$, in the total observing time, $t_{\rm obs}$, over a bandwidth, $B_{\rm obs}$, is given by
\begin{equation}
N_{\rm scint} = \left(\frac{B_{\rm obs}t_{\rm obs}}{\nud \td}\right)f_d,
\label{eq:Nscint}
\end{equation}
where $f_d$ is a filling fraction that describes how much of the observed frequency-time phase space contains signal \citep[e.g.][]{1999ApJSBhat}, which we assume to be $f_d \approx 0.5$ based on the Parkes dynamic spectrum in Figure~\ref{fig:dyn_spec}.
For the MWA, where the scintles are on the order of 10\,kHz wide, this factor is negligible ($N_{\rm scint}\approx 2\times 10^5$ and $\sigma_{\rm stat} \sim 0.2\%$, or 20\,Hz).
However, in the case of Parkes, we clearly sample far fewer scintles ($N_{\rm scint}\approx 2$), ergo the sampling error is $ \sigma_{\rm stat} \approx 70\%$ and $\nud = \nudPKS$\,MHz (where the final error is the quadrature sum of the fitting error and statistical error).

\subsubsection{Scintillation time scale}
\begin{figure*}
\centering
\includegraphics[width=\textwidth]{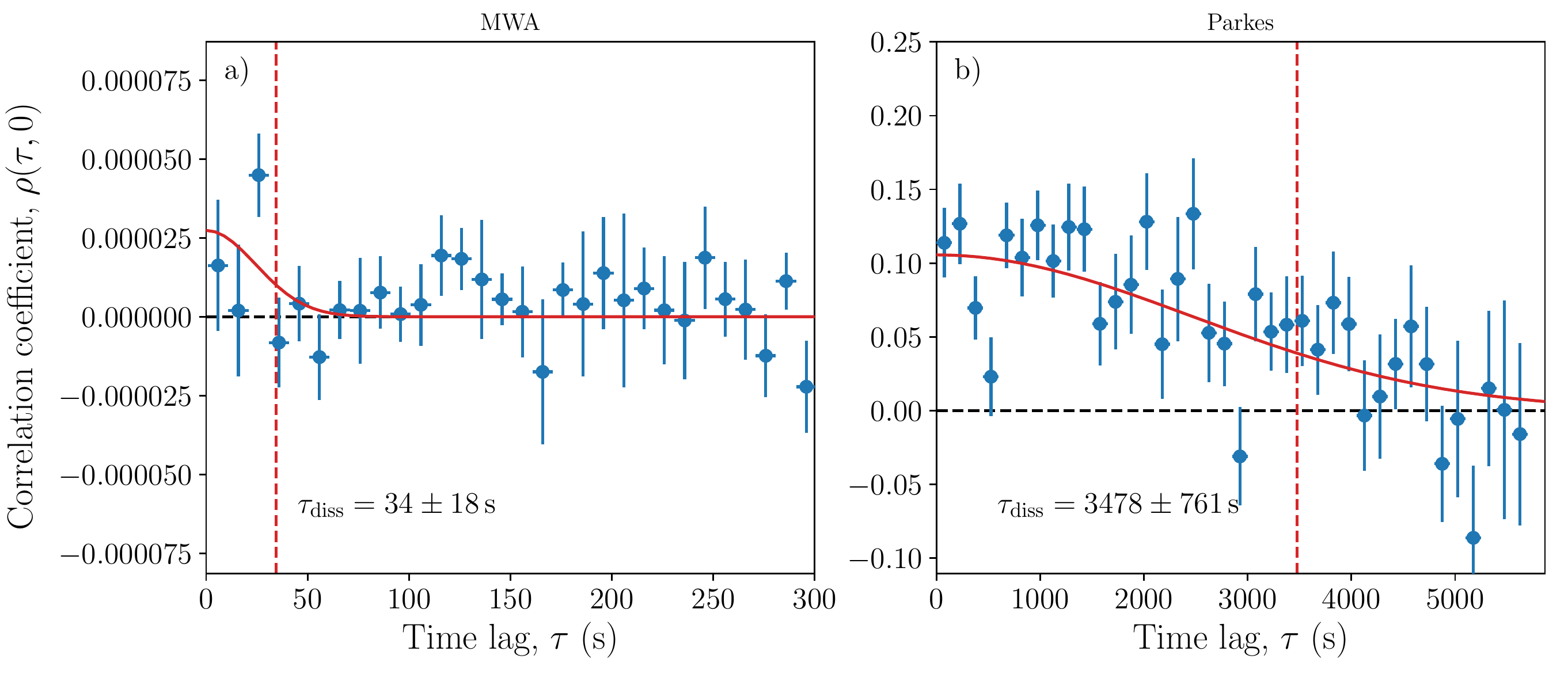}
\caption{Mean correlation coefficients of individual pulse spectra, binned into: a) 10-second intervals for MWA data, and b) 150-second intervals for Parkes data.
The Gaussian fit to the data (red, solid line) is weighted based on the standard error of each of the points, where the $1/e$-half-width of the Gaussian corresponds to the scintillation time scale.
We measure a scintillation time scale $\td=\tdPKS$\,s at 1.4\,GHz, and $\td=\tdMWA$\,s at 154\,MHz, marked by the vertical dashed red lines.}
\label{fig:mwa_pks_tcorr}
\end{figure*}

The RRAT emission irregularly samples the scintillation pattern which makes estimating the scintillation time scale, $\td$, difficult. 
Nevertheless, we calculate the intensity cross-correlation,
\begin{equation}
\rho(\tau, \delta\nu=0)=\langle \Delta I(t, \nu) \Delta I(t+\tau, \nu+\delta\nu)\rangle
\end{equation}
between every mean-subtracted single-pulse spectra, $\Delta I(t,\nu)$, and subsequent pulses, while recording the corresponding time lag, $\tau$, as the number of pulsar rotations between the correlated pulse spectra\footnote{The actual correlation is implemented using the NumPy function {\tt numpy.correlate} with {\tt mode=same}.}.
We average the correlation coefficients for each time lag and then re-bin the results such that there is one $\rho(\tau,0)$ per 150 seconds for the Parkes data.
In Figure~\ref{fig:mwa_pks_tcorr} we plot these correlation coefficients against time lag.

The scintillation time scale is the $1/e$-half-width of the fitted Gaussian ($\td=\sqrt{2}\sigma$), where the mean is forced to zero, which we measure to be $\td = 3478\pm 761$\,s at 1.4\,GHz.
Including the relative sampling error of $\sim 70\%$, we find that $\td = \tdPKS$\,s.
Additionally, we can estimate the expected refractive interstellar scintillation (RISS) time scale \citep[e.g.][]{1990ARA&ARickett} at 1.4\,GHz, where $\tau_{\rm riss} = \td\left(\nu/\nud\right) \approx 13$\,hours.
These values should be considered with caution given that, as for the scintillation bandwidth, we do not actually sample a full scintle over the 1.5\,hour observation.

Scaling the decorrelation time from 1.4\,GHz, assuming $\td\propto \nu^{1.2}$, implies $\td\approx 250\pm 185$\,s at 154\,MHz.
Using the same technique as above on the MWA data, except re-binning to one $\rho(\tau,0)$ per 10 seconds (given the expected $\td$), we find that we are severely limited by the signal-to-noise ratio of our detected pulses.
This is due to the scintillation bandwidth being on the order of, or less than, the channel width.
The estimated scintillation time scale is a factor of 7 less than expected, where $\td = 34\pm 18$\,s.
Visually inspecting Figure~\ref{fig:mwa_pks_tcorr} one can see that the correlation coefficients, even after averaging in time, are consistent with noise except for one outlier. 
Furthermore, the quality of the Gaussian fit changes drastically depending on how the correlation coefficients are averaged, and generally results in an unconstrained estimate of $\td$ (i.e. undefined uncertainty or failing to find an adequate fit altogether).
For these reasons we caution against interpretation of the measured scintillation parameters at 154\,MHz alone.

\subsubsection{Scintillation velocity and turbulence strength}
We can calculate the scintillation velocity---as a proxy to the RRAT space velocity---under the caveats that: both $\nud$ and $\td$ are nominally lower limits; that we do not know the relative distance of the scattering screen to the pulsar, and; that there is a factor of 2 discrepancy in distance estimates, where $D=0.7$\,kpc and $D=1.49$\,kpc, from NE2001 \citep{2002ArXivCordes,2003ArXivCordes} and YMW2016 \citep{2017ApJYao} models, respectively.
We assign a 25\% uncertainty to each distance estimate.
The scintillation velocity is given by
\begin{equation}
V_{\rm iss} = A_{\rm iss} \frac{\left(Dx\nud\right)^{1/2}}{\nu\td},
\label{eq:viss}
\end{equation}
where $D = D_{\rm os} + D_{\rm ps}$ is the total distance to the pulsar in kpc; $D_{\rm os}$ and $D_{\rm ps}$ are the distances from the screen to the observer and pulsar respectively in kpc; $x = D_{\rm os} / D_{\rm ps}$ (in this case we assume $x=1$, i.e. the screen is located exactly half way between the observer and the pulsar); $\nu$ is the observing frequency in GHz; $\nud$ is in MHz, and $\td$ in seconds.
The scaling constant $A_{\rm iss}=2.53\times 10^4\,{\rm km\,s^{-1}}$ is derived for a homogeneously distributed ISM with a Kolmogorov turbulence spectrum \citep{1998ApJCordes}, which appears to be a valid approximation for this pulsar given the scintillation frequency scaling index calculated in Section~\ref{sec:scint_bw}.
For a distance $D=0.70\pm0.18$\,kpc we find $V_{\rm iss}=\vissNE\,{\rm km\,s^{-1}}$, while for $D=1.49\pm0.37$\,kpc, $V_{\rm iss}=\vissYMW\,{\rm km\,s^{-1}}$.
The uncertainties correspond to the quadrature sum of the scintillation bandwidth, scintillation time scale and distance errors, given by
\begin{multline}
\Delta V_{\rm iss} = \left(
\left[\frac{\partial V_{\rm iss}}{\partial \nud}\Delta\nud\right]^2 \right. + \\
\left[\frac{\partial V_{\rm iss}}{\partial \td}\Delta\td\right]^2 +
\left. \left[\frac{\partial V_{\rm iss}}{\partial D}\Delta D\right]^2 \right)^{1/2},
\end{multline}
where $\Delta X$ represents the uncertainty in parameter $X$. 
We did not calculate the scintillation velocities from the MWA data since we do not have reliable estimates of $\nud$ and $\td$ (see Table~\ref{tab:scint_properties}).

We can also place limits on the mean turbulence strength, $\cn$, along the line-of-sight.
Assuming Kolmogorov turbulence, the mean turbulence strength in units of ${\rm m^{-20/3}}$ is
\begin{equation}
\cn \approx 0.002\,\nu^{11/3} D^{-11/6} \nud^{-5/6},
\end{equation}
(cf. eq.~9 of \citealp{1990ApJCordes}) where $\nu$, $D$ and $\nud$ are in the same units as for eq.~\eqref{eq:viss}.
Given the range in distances, we find that the corresponding range in turbulence strength is $\cn\approx (7\text{--}28)\times 10^{-5}\,{\rm m^{-20/3}}$ at 1.4\,GHz, and note that given the lower limit on $\nud$ from Parkes data we can only confidently say that $\cn \lesssim 2.8\times 10^{-4}\,{\rm m^{-20/3}}$.

\subsection{Spectral index}\label{sec:spec_index}
\begin{figure}
\centering
\includegraphics[width=\columnwidth]{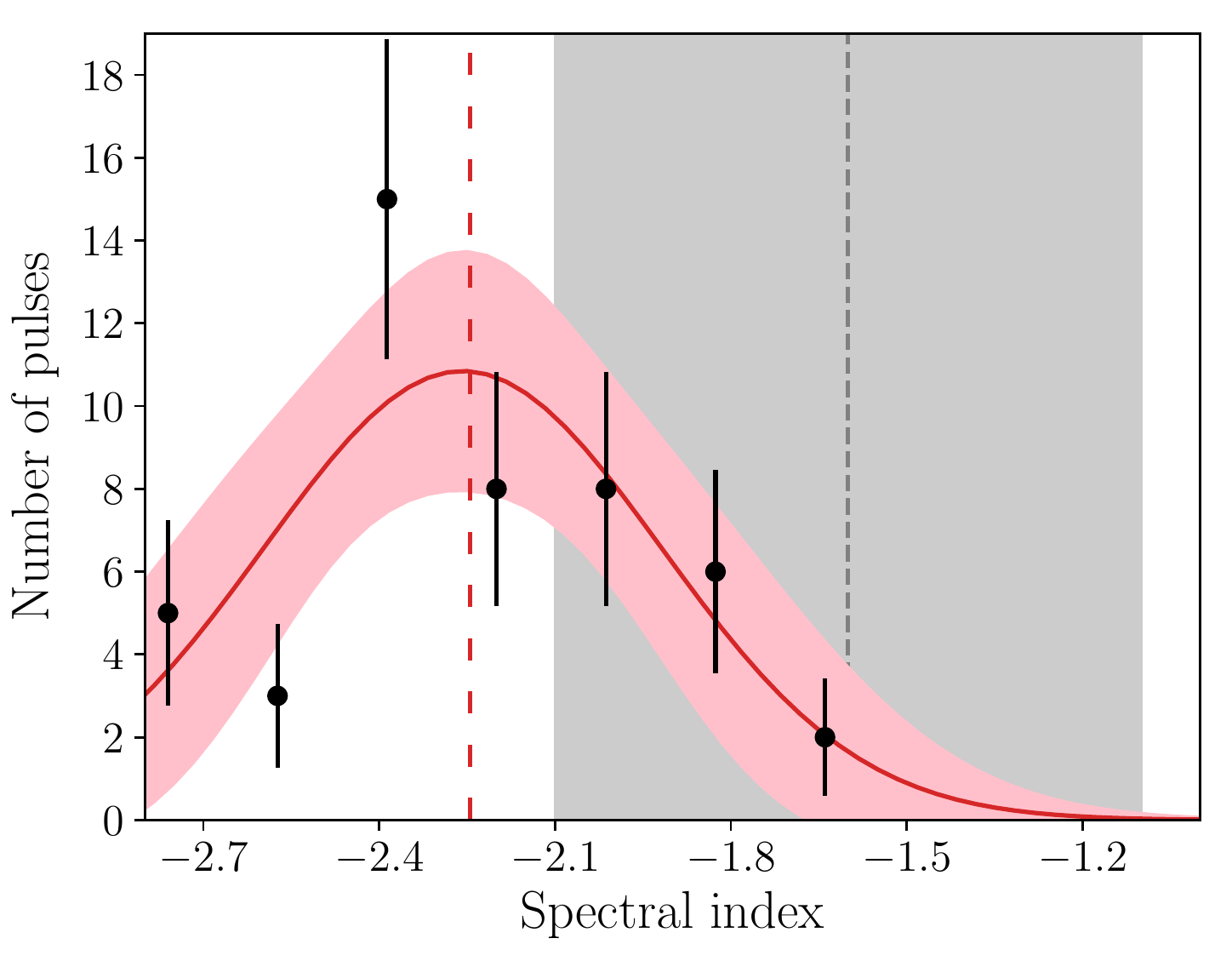}
\caption{Spectral index distribution for detected simultaneous pulses between the MWA and Parkes.
The red solid line is a Gaussian fit to the distribution, and the pink envelope represents the 1-$\sigma$ confidence interval of the model.
Error bars on the points are Poisson uncertainties only.
The mean spectral index is $\alpha=-2.2\pm 0.1$ with a standard deviation of $\sigma=0.4\pm 0.1$.
The grey shaded region is the typical distribution of spectral indices, with a mean of $\langle\alpha\rangle=-1.6$ and standard deviation of $\sigma=0.5$ \citep{2018MNRASJankowski}. }
\label{fig:alphas_dist}
\end{figure}

A cross-matched list of single-pulses was created from the independent MWA and Parkes pulse data sets using the STILTS table manipulation software \citep{2006ASPCTaylor}.
After the cross-matching stage, there were \nMatch\ pulses coincident in both bands.
For each of these pulses we used their measured fluences to calculate a spectral index, the distribution of which is shown in Figure~\ref{fig:alphas_dist} along with an indicator of the normal pulsar population spectral index range in grey.

We measure a mean single pulse spectral index of $\alpha_{154}^{1369}=-2.2\pm 0.1$ which is relatively steep compared to mean spectral index observed in the typical pulsar population, where $\langle\alpha\rangle\approx -1.6$ (see e.g. \citealp{2000A&ASMaron,2013MNRASBates,2018MNRASJankowski}).
The range of single-pulse spectral indices we measure is $-2.8 < \alpha_{154}^{1369} < -1.5$.
The steep spectral index we measure seems to agree empirically with the detections reported in the literature, given that RRAT \rrat\ has been detected multiple times with low-frequency observations in the past from the GBT (350\,MHz), LOFAR (150\,MHz) and LWA1 (30--80\,MHz).

\subsection{Fluence distributions}\label{sec:fluence_dists}
\begin{figure*}
\centering
\includegraphics[width=\textwidth]{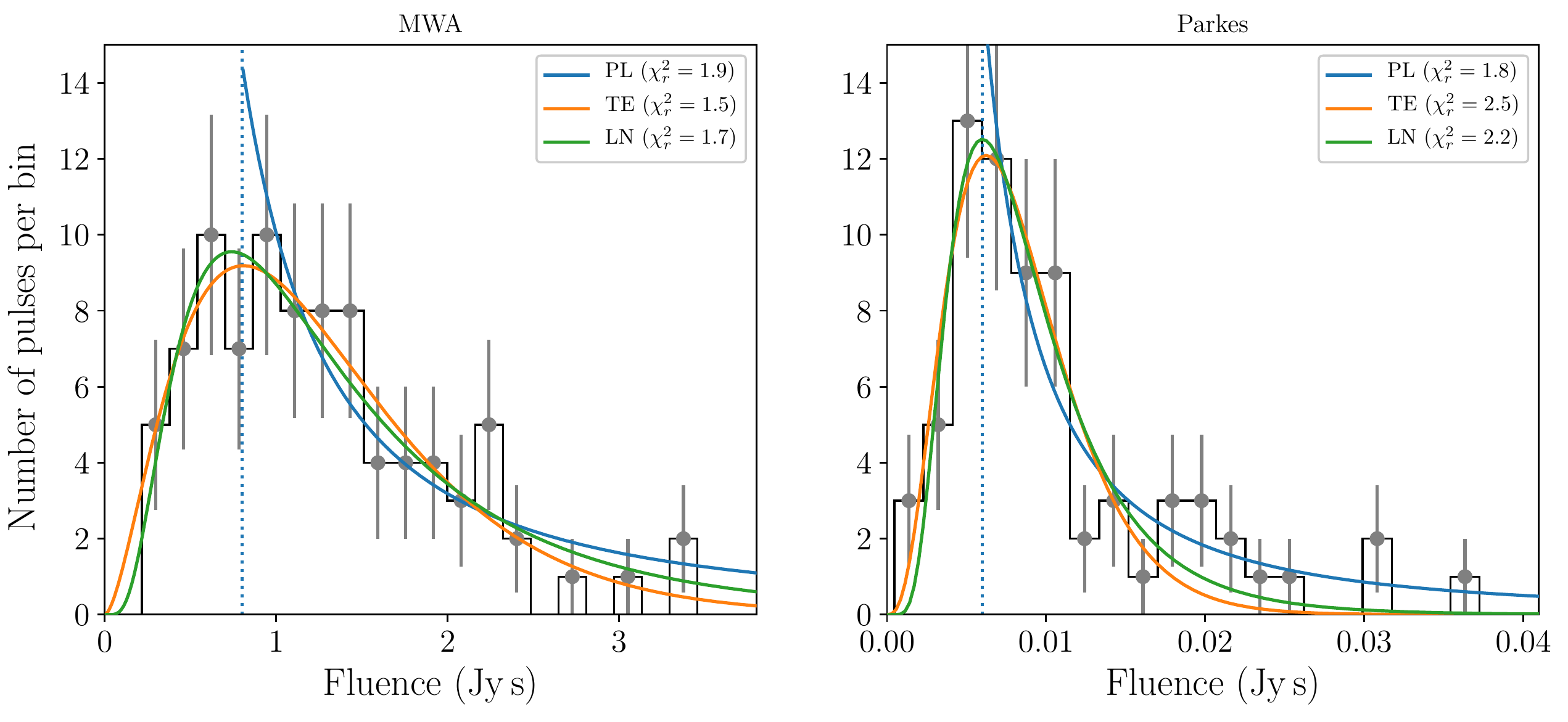}
\caption{Pulse fluence (energy) distributions for single pulses detected with the MWA (left) and Parkes (right).
We fitted a power law (blue), truncated exponential (orange) and log-normal (green) distribution model to the binned single-pulse fluences. 
The error bars represent statistical (Poisson) errors only.
The reduced chi-squared values of the fits are given in the legend. 
The power law cut-off for each frequency is indicated by the blue vertical dotted lines.
}
\label{fig:fluence_dist}
\end{figure*}

From the detected pulses we constructed fluence (pulse energy) distributions for each band. 
To these distributions we fit three relativity common models using the Python \textsc{lmfit} module\footnote{\url{https://github.com/lmfit/lmfit-py} (v0.9.11, doi: 10.5281/zenodo.1301254)}: a power law (PL), a truncated exponential (TE; functionally the same as eq.~3  of \citealp{2018MNRASMickaliger}), and a log-normal distribution (LN).
The relevant functional forms are:
\begin{align}
N_{\rm PL}(x) &= Ax^{-\beta},\\
N_{\rm TE}(x) &= Bx^{-\zeta}e^{-\lambda x},\\
N_{\rm LN}(x) &= \frac{C}{x\sigma}\exp\left[-\frac{(\ln x -\mu)^2}{2\sigma^2}\right],
\end{align}
where $N(x)\,{\rm d}x$ is the number of pulses at fluence $x$, $A$, $B$ and $C$ are arbitrary scaling constants, $\beta$ and $\zeta$ are power law exponents, $\lambda$ is a decay parameter, and $\mu$ and $\sigma$ are the location and scale parameters for the normally distributed logarithm (i.e. $\ln x)$.
Note that in this case the power laws are fitted only to the pulses which have fluences greater than 0.8 and 0.006\,Jy\,s for the MWA and Parkes, respectively.
These cutoffs were chosen to coincide roughly with where the distributions peak.
Without these restrictions, the power law model is a poor fit to the data.
The data and fitted models are shown in Figure~\ref{fig:fluence_dist}, and model parameters (with standard errors) are given in Table~\ref{tab:fluence_best_fit}.
In general it appears that a log-normal distribution is favoured, though see Section~\ref{sec:disc_fluences} for further discussion.

\begin{table*}
\caption{Best-fit parameters for trial fluence distribution models. \label{tab:fluence_best_fit}}
\centering
\begin{tabular}{@{}ccccccccc@{}}
\hline\hline
 & \multicolumn{2}{c}{Power law$^{*}$} & \multicolumn{3}{c}{Truncated exponential} & \multicolumn{3}{c}{Log-normal}\\
Frequency & $\beta$ & $\chi_r^2$ & $\zeta$ & $\lambda$ & $\chi_r^2$ & $\mu^{a}$ & $\sigma^{a}$ & $\chi_r^2$\\
(MHz) & & & & ($\rm Jy^{-1}\,s^{-1}$) & & ($\rm Jy\,s$) & ($\rm Jy\,s$) & \\
\hline
154 & $1.6\pm0.2$ & 1.9 & $-1.7\pm0.4$ & $2.1\pm0.3$ & 1.5 & $0.18\pm0.07$ & $0.69\pm0.06$ & 1.7\\
1369 & $2.2\pm0.4$ & 1.8 & $-3\pm0.7$ & $488\pm97$ & 2.5 & $-4.8\pm0.1$ & $0.53\pm0.05$ & 2.2\\
\hline\hline
\end{tabular}
\medskip
\tabnote{$^*$Restricted to fitting pulses with fluences greater than 0.8 and 0.06\,Jy\,s for the MWA and Parkes (see text).}
\tabnote{$^a$The location ($\mu$) and scale ($\sigma$) parameters, as defined by Python's {\tt scipy.stats.lognorm}.}
\end{table*}

\subsection{Pulse rates and clustering}\label{sec:pulse_rates}
We measured a total of \nMWA\ and \nPKS\ pulses with $\rm{S/N}\geq 6$ with the MWA and Parkes, respectively.
These detections correspond to pulse rates of $73\pm 7\,{\rm hr^{-1}}$ above a peak flux density of 65\,Jy at 154\,MHz, and $43\pm 5\,{\rm hr^{-1}}$ above a peak flux density of 0.6\,Jy at 1.4\,GHz, where the uncertainties correspond to the Poisson counting error.
The pulse rates we measure, as well as those in the literature are presented in Table~\ref{tab:pulse_rates}.
In the case of the minimum detectable flux density for the LOFAR results \citep{2015ApJKarako-Argaman}, we assign a gain for the core stations of $0.68\mathrm{\,K\,Jy^{-1}}$ (based on estimates of the collecting area of \citet{2013A&AvanHaarlem}, modified by a projection factor of $\cos^2(\pi/6)=3/4$ assuming a best-case scenario where the source was observed at $\sim 60$\,deg elevation) and add 250\,K to the nominal 400\,K receiver temperature in an attempt to include sky noise contributions.
Using these values, we estimate a minimum detectable flux density of $\sim 21$\,Jy, assuming the same caveats of the original estimate (i.e. $\text{SNR}\geq 5 $ and 10\,ms pulse width).

\begin{table*}
\caption{Pulse rates and nominal detection sensitivity for single-pulses from RRAT \rrat. \label{tab:pulse_rates}}
\centering
\begin{tabular}{@{}cccccccc@{}}
\hline\hline
Telescope & Frequency & Bandwidth & Min. $S_\nu$ & Min. fluence$^a$ & Observing time & Pulse rate & Ref.\\
          & (MHz) & (MHz) & (Jy) & ($\rm Jy\,s$) & (hours) & (hr$^{-1}$) & \\
\hline
LWA1 & 35.1, 49.8, 64.5, 79.2 & $4\times 15$ & $\sim 60^{b}$ & $\sim 0.3$ & 26$^{c}$ & 12--21 & T+16\\
LOFAR & 150 & 80 & $\sim 21^{d}$ & $\sim 0.11$ & 0.75 & $52\pm 8$ & KA+15\\
MWA & 154 & 30.72 & $\sim 65$ & 0.19 & 1.4 & $73\pm 7$ & \textit{This work}\\
GBT & 350 & 100 & $\sim 0.4^{d}$ & $\sim 0.002$ & $\sim 0.5$ & $46\pm 9$ & KA+15\\
Parkes & 1369 & 256 & $\sim 0.6$ & 0.0005 & 1.6 & $43\pm 5$ & \textit{This work}\\
\hline\hline
\end{tabular}
\medskip
\tabnote{References --- T+16: \citet{2016ApJTaylor}, KA+15: \citet{2015ApJKarako-Argaman}.}
\tabnote{$^{a}$Fluence limits from other works were estimated by calculating the area under a tophat with amplitude equal to the corresponding sensitivity and a width of 5\,ms (mean effective width of pulses measured in this work).}
\tabnote{$^{b}$Over the full observed bandwidth, which was at best 60\,MHz. Individual subband sensitivities are therefore $\sim 120$\,Jy.}
\tabnote{$^{c}$Split into 2--3 hour blocks over 10 days spread throughout late-2013 and late-2014, see \citet{2016ApJTaylor} for details.}
\tabnote{$^{d}$Calculated using eq.~1 and observing parameters from Table~1 (though see text regarding LOFAR parameters) of \citet{2015ApJKarako-Argaman}, assuming a detection threshold of ${\rm S/N}\geq 5$.}
\end{table*}

\begin{figure}
\centering
\includegraphics[width=\linewidth]{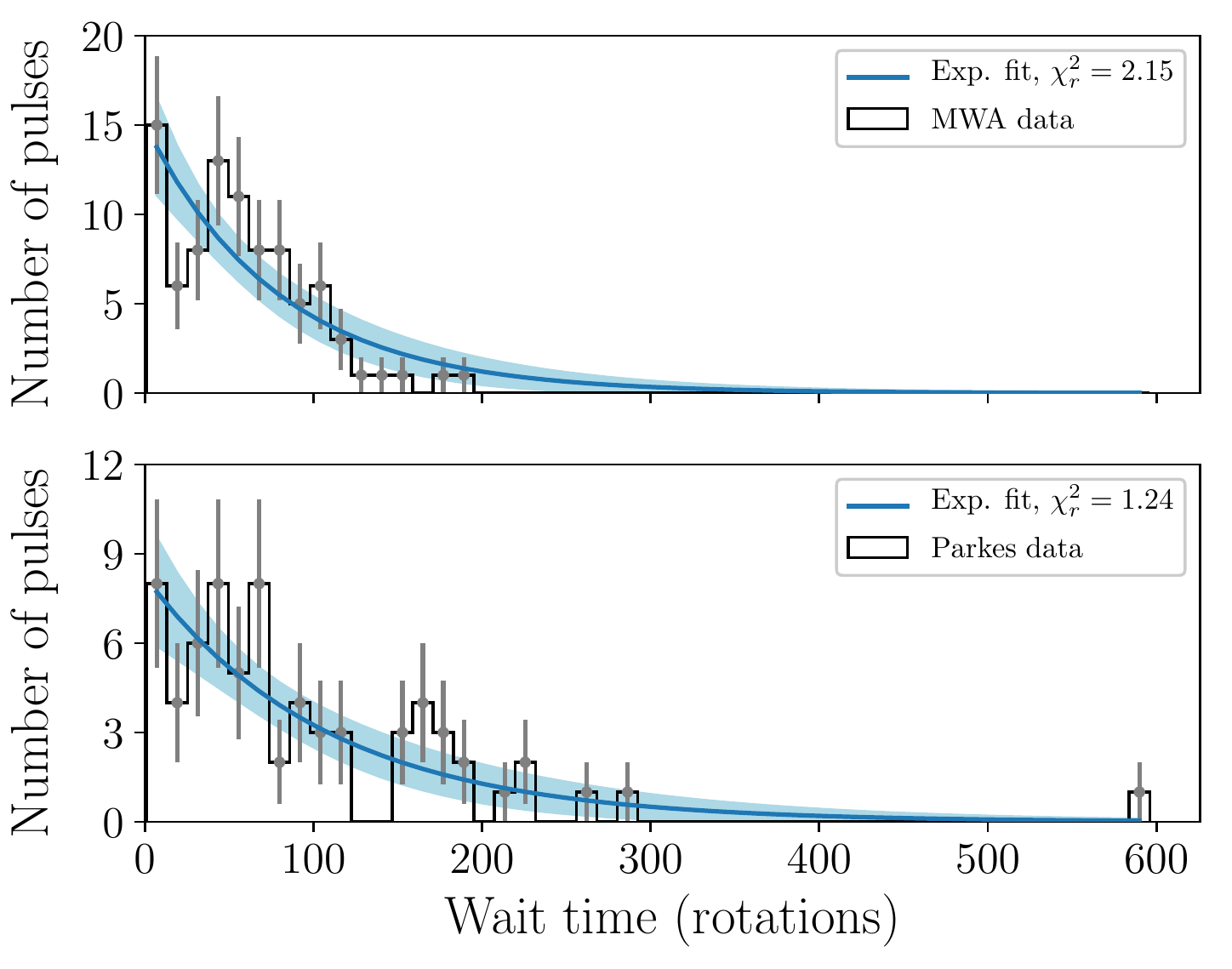}
\caption{Histogram of the number of rotations between subsequent pulses (i.e. wait times) for the MWA pulses (top) and Parkes pulses (bottom). 
The blue solid lines are a fit to an exponential distribution, and the light blue shaded regions represent the 99\% confidence interval based on the fitting uncertainties.
Wait times were binned into 50 equally spaced bins ranging from 1 rotation to 596 rotations (i.e. 517 seconds, the maximum wait time in either frequency band).
The reduced chi-square statistic, $\chi_r^2$, for each fit is 2.15 and 1.24 for the MWA and Parkes, respectively.
}
\label{fig:wait_time}
\end{figure}

We also examine the distribution of the number of rotations between subsequent pulses (``wait times'') within our observation.
These wait times are presented in Figure~\ref{fig:wait_time}.
In this case, we binned the wait times into 50 equally spaced intervals, ranging from one pulsar rotation to the maximum wait time, which corresponds to 596 rotations ($\sim 517$\,s) in the Parkes data. 
The median wait times for the MWA and Parkes pulses are 52 and 68 rotations, respectively.
In both the MWA and Parkes data, the minimum wait time is one rotation.
The maximum wait time in the MWA data is 184 rotations (i.e. $\sim 159$ seconds), with 75 instances of wait times less than 100 rotations.
For Parkes, the maximum wait time is substantially longer, at 596 rotations (i.e. $\sim 520$ seconds), and there are 46 wait times less than 100 rotations.
Given the different sensitivity thresholds of each telescope, it is difficult to quantitatively compare these numbers, especially since scaling thresholds and selecting pulses from either sample only adds to the issue of small number statistics in this case.

If the single-pulse emission is produced by a Poisson (random) process, then we would expect that the time between events (i.e. the wait times) would be exponentially distributed ($N(x)\,{\rm d}x \propto e^{-\eta x}$).
After fitting each sample independently, we find that the exponents are similar, where $\eta_{\rm MWA} = 0.013\pm 0.001$ for the MWA, and $\eta_{\rm PKS} = 0.009\pm 0.001$ for the Parkes data.
From these data, it is unclear whether there is a significant excess beyond what would be expected of pulse events drawn from a Poisson distributed process (an exponential distribution has been fitted to the wait times, see Figure~\ref{fig:wait_time}).
In the context of the general pulsar population, similar work has be done for nulling pulsars, noting examples of clustering \citep[e.g.][]{2009MNRASRedman}, and of random processes \citep[e.g.][]{2012MNRASGajjar}.
The latter is reasonably consistent with what we find for RRAT \rrat.
Ultimately, we are limited in this case by the number of single-pulse detections, and note that the Parkes wait time distribution will be biased by the scintillation effects.

\subsection{Rotation measure}\label{sec:rm}
The rotation measure (RM) quantifies the degree of Faraday rotation that the radio emission from a source experiences after traversing the ISM, and was traditionally measured by calculating the change in linear polarisation angle across the observing band \citep[e.g.][]{2008MNRASNoutsos,2018ApJSHan}.
We used the PSRCHIVE rotation measure fitting routine, \textsc{rmfit}, which effectively implements the RM synthesis method \citep[e.g.][]{2005A&ABrentjens}, to determine the nominal rotation measure of the Parkes and MWA data based on the polarisation properties of the pseudo-integrated pulse profiles (Figure~\ref{fig:profiles}).

The ionosphere can significantly contribute to the measured RM, thus we calculated the ionospheric contribution for both Parkes and the MWA using \textsc{ionFR}\footnote{\url{http://ascl.net/1303.022}} \citep{2013A&ASotomayor-Beltran}, using the latest version of the International Geomagnetic Reference Field \citep[IGRF12;][]{2015EP&SThebault} and the International Global Navigation Satellite System Service vertical total electron content maps \citep[e.g.][]{2009JGeodHernandez-Pajares}. 
The RM contribution from the ISM is given by $\rm RM_{ISM} = RM_{obs} - RM_{ion}$, the values for which are given in Table~\ref{tab:rm}.
The ionosphere was relatively quiet during the observations, but is still the dominant source of uncertainty in estimating the RM imparted by the ISM for the MWA measurements.

\begin{table}
\caption{Rotation measure estimate for RRAT \rrat.}
\centering
\begin{tabular}{@{}cccc@{}}
\hline\hline
Frequency & $\rm RM_{obs}$ & $\rm RM_{ion}$ & $\rm RM_{ISM}$ \\
(MHz)     & ($\rm rad\,m^{-2}$) & ($\rm rad\,m^{-2}$) & ($\rm rad\,m^{-2}$) \\
\hline
154 & $3.38\pm 0.03$ & $-0.47\pm0.09$ & $3.85\pm 0.12$\\
1396 & $2.41\pm 2.74$ & $-0.41\pm 0.15$ & $2.82\pm 2.89$\\
\hline\hline
\end{tabular}
\label{tab:rm}
\end{table}

After ionospheric correction, the ISM contribution to the RM along this line-of-sight based on the MWA data, is $\rm RM_{ISM}=3.85\pm 0.12\,rad\,m^{-2}$.
While the Parkes data measurement is less constraining, it does agree within uncertainty.
Given the RM and DM we can estimate the average line-of-sight magnetic field strength using the approximation
\begin{equation}
\langle B_\parallel\rangle \approx 1.23\left(\frac{\rm RM}{\rm DM}\right) \mathrm{\,\mu G},
\end{equation}
where we find that $\langle B_\parallel\rangle\approx 0.32\pm 0.01\mathrm{\,\mu G}$, where the uncertainty is the quadrature sum of the relative error in the DM and RM measurements.
While reasonably small, this magnetic field strength is well within the distribution of measured values for larger samples of pulsars over a wide range of Galactic latitudes \citep[e.g.][]{2003A&AMitra,2008MNRASNoutsos,2019MNRASSobey}.

\section{DISCUSSION}\label{sec:discussion}
\subsection{Scintillation characteristics compared to normal pulsars}\label{sec:disc_scint}
The analysis we present is the first direct example of detected scintillation from RRATs.
Generally, parameterising the scintillation can characterise the turbulence in the ISM and estimate pulsar space velocities.
While scintillation is expected for these objects, it is technically difficult given that the sporadic nature of RRAT emission will often hinder the robust characterisation of parameters.
In particular, the irregular sampling of the intensity modulation in time makes estimating the scintillation time scale more difficult, while in our case we are also limited by the bandwidth (in the case of the 1.4\,GHz data) and frequency resolution (in the case of the 154\,MHz data).
Nonetheless, we have attempted to constrain the scintillation bandwidth and time scale (and related quantities) for RRAT \rrat\ based on our observations of $\sim 100$ single pulses over $\sim 5800$\,s.

For RRAT \rrat, the full scintle size (in frequency) at 1.4\,GHz is considerably larger than the 256\,MHz observing bandwidth; thus, we interpret the measured scintillation bandwidth of $\nud=\nudPKS$\,MHz as a lower limit.
The predicted scintillation bandwidth from NE2001 along the line-of-sight to RRAT \rrat\ is $\nud^{\rm NE2001} = 27^{+20}_{-9}$\,MHz which is a factor of $\sim 4$ lower than what we measure.
This is not necessarily alarming given that the NE2001 model attempts to model the turbulence within the ISM, largely based on Galactic plane measurements, thus a factor of a few discrepancy is expected for objects with large Galactic latitudes.
Furthermore, it is known that the measured scintillation properties of nearby pulsars are modulated by factors of $\sim 3\text{--}5$ over time \citep{1999ApJSBhat,2016ApJLevin}.

The scattering strength, $u=(\nu/\nud)^{1/2}\approx 4$, suggests that, at 1.4\,GHz, we are in the strong scintillation regime ($u > 1$).
It also implies that we are sampling only a small range of turbulence scale sizes in the ISM.
This is consistent with the calculated turbulence strength (Table~\ref{tab:scint_properties}) and with expectations based on the Galactic latitude of the pulsar ($b=-60.2^\circ$).
The turbulence towards RRAT \rrat\ is typical of nearby pulsars, especially when comparing the turbulence strength we calculate, $\cn \lesssim 2.8\times 10^{-4}\,{\rm m^{-20/3}}$, to other pulsars with anomalously reduced turbulence.
For example, the ISM along the line-of-sight to PSR J0437$-$4715 is, on average, $\sim 30\%$ as turbulent as towards RRAT \rrat\ ($\cn = 8\times 10^{-5}\,{\rm m^{-20/3}}$; \citealt{2018ApJSBhat}), and the ISM towards PSR J0953+0755 is only $\sim 7\%$ as turbulent ($\cn \sim 2\times 10^{-5}\,{\rm m^{-20/3}}$; \citealt{1992NatPhillips}).
Overall, the scintillation properties\footnote{An important note regarding these estimates is that we are fundamentally biased, because even if over the observation more than one full scintle passes through the line-of-sight, we do not fully sample it given that we are only counting the ${\rm S/N}\geq 6$ pulses (i.e. if some pulses are scintillating down then we simply don't detect them in this collection).} of RRAT \rrat\ suggest that this is a relatively typical line-of-sight through the ISM.

The scintillation velocity for RRAT \rrat\ is relatively small, though is within the measured range for normal pulsars \citep[e.g.][]{1982NaturLyne,1986ApJCordes,1998MNRASJohnston} with similarly low/moderate DMs.
This further supports the idea that RRAT \rrat\, and the ISM surrounding it is not particularly anomalous compared to normal pulsars or other sight-lines.
We have shown that scintillation analysis is a feasible way of constraining the space velocities of RRATs, and placing them in context with the broader pulsar population.
This kind of analysis could be particularly powerful when making use of both wideband receiver backends and long duration observations, especially with next generation instruments.

At 154\,MHz the scintles are too small to robustly resolve, thus our estimates of $\nud\lesssim 10$\,kHz and $\td = \tdMWA$\,s should be interpreted with caution.
An intermediate frequency band, in the range of 300--700\,MHz, would be ideal to further characterise the scintillation properties of this pulsar.
At 500\,MHz, assuming $\nud\propto\nu^{-4.2}$ and scaling from the Parkes values, one would expect that the scintillation bandwidth is $\nud\approx 1.4$\,MHz, which should be resolvable with careful selection of observing parameters.
Furthermore, observing the pulsar over a very wide bandwidth (e.g. the newly commissioned ultra-wide bandwidth receiver at Parkes, which samples 0.7--4\,GHz contiguously) would enable us to resolve $>1$ scintles and therefore more robustly estimate the scintillation parameters, even though the steep spectral index could impact the detections at higher frequencies.

\subsection{Spectral index}\label{sec:disc_spec_index}
The spectral index distribution of RRATs is poorly explored.
Recently \citet{2018ApJShapiro-Albert} provided single-pulse based spectral index measurements (similar to those presented here) for three RRATs (J1819$-$1458, J1913+1313 and J1317$-$5759) and find that while the average spectral indices fall within the normal pulsar population distribution, they are typically flatter than normal pulsars, which is in contradiction to what we find for RRAT \rrat. 
That being said, given that the mean spectral index measurement is based only on simultaneously detected pulses, our analysis is subject to a selection effect, whereby not including shallower single-pulse spectral indicies simply because they are not detected with the MWA could act to artificially steepen the measured value.
The authors also note a wide range of single-pulse spectral indices ($-7<\alpha<+4$), which could be due to the intrinsic emission process or because the spectral indices were measured within the observed bandwidth (where the frequency lever-arm is relatively small).
Spectral indices are known to widely vary from pulse-to-pulse, for normal pulsars \citep[e.g.][]{2003A&AKramer}, and giant pulses from the Crab pulsar \citep[e.g.][]{2010A&AKaruppusamy,2017ApJMeyers}; so, it is not necessarily surprising that this is also true for RRATs.

The steep spectral index we find for RRAT \rrat\ indicates that at least some RRATs also exhibit a relatively large range of average spectral indices, just like the normal pulsar population.
We note though that the scintillation occurring at 1.4\,GHz will bias the mean spectral index to be steeper than in reality (assuming no intrinsic time variability in the emission).
\citet{2016ApJTaylor} measure a spectral index $\alpha_{35}^{80} \approx -0.7$ across the observing band of LWA1 (35--80\,MHz), which is substantially shallower than what we measure (though consistent with the trend observed by \citealp{2015ApJStovall}) and indicates that a spectral flattening may occur at frequencies $\lesssim 150$\,MHz.
From a detection and characterisation perspective, this means that the next generation of radio telescopes (e.g. the SKA and ngVLA) and their pulsar/fast-transient search experiments will be in a position to compliment each other, where the low and high-frequency observations will together sample a wider range of (and possibly different) objects in the pulsar population.

\subsection{Fluence distributions}\label{sec:disc_fluences}
Our ability to robustly distinguish between pulse energy distribution models is somewhat diminished given the relatively small number of pulses available.
As previously noted, the power law models should be interpreted carefully as they only provide reasonable fits to pulses above an arbitrary fluence cut-off. 
This is an unfortunate, and often ignored, bias that is difficult to correct even when using more complicated fitting methods and larger data sets.
Given that the data being used are only real pulse detections, any model should nominally be able to account for the high and low-energy pulses simultaneously, and the power law cannot do this.
Nevertheless, comparing the relative reduced chi-squared statistics, it seems that the truncated exponential and log-normal distributions are similarly good fits to the data, while the power law is nominally a better match if we ignore the caveat of arbitrary fluence cut-offs. 

Given that the normal pulsar population is seen to exhibit mostly log-normal pulse energy distributions \citep[e.g.][]{2012MNRASBurke-Spolaor} and that there has been recent work showing similar results for RRATs and intermittent pulsars \citep[e.g.][]{2017ApJCui,2018MNRASMickaliger,2018ApJShapiro-Albert,2018ApJMeyers}, it appears that RRAT \rrat\ follows the trend. 
An important caveat is that comparing amplitude (i.e. peak flux density) distributions to fluence (i.e. pulse energy) distributions can be confusing given that the distribution types and parameters do not necessarily map one-to-one.
In our case though, we are unable to unequivocally state that a log-normal model is the best fit without further observations and a larger sample of single pulses.
Furthermore, a log-normal distribution does seem to align with the evidence in the literature. 
Ignoring the fact that an arbitrary fluence cut-off was employed, the power law indices measured, at least for the Parkes data, are similar to those seen for Crab giant pulses \citep[e.g.][]{2008ApJBhat,2012ApJMickaliger}, though it is difficult to confidently determine whether there is actually a steep power law tail, given our small number of detections.
We do not see evidence of multiple overlapping distributions for RRAT \rrat\ pulse energies, though, again, a larger sample of pulses is required to robustly test this.

\subsection{Pulse rates and clustering}\label{sec:disc_rates}
The majority of RRATs do not have published pulse rates making it difficult to compare our measured rates for RRAT \rrat\ to the overall population.
For \rrat, the previously published pulse rates are: $46\pm 9\,{\rm hr^{-1}}$  with the GBT at 350\,MHz, $52\pm 8\,{\rm hr^{-1}}$ with LOFAR at 150\,MHz \citep{2015ApJKarako-Argaman}, and $12\text{--}21\,{\rm hr^{-1}}$ with the LWA1 between 35--79\,MHz \citep{2016ApJTaylor}.
Our measured pulse rate from the Parkes data ($43\pm 5\,{\rm hr^{-1}}$) is in relative agreement with these values, though we again note the importance of considering the effects of scintillation, and a different sensitivity threshold per instrument, on the detection statistics.
From the perspective of the MWA, we find that the pulse rate is somewhat higher ($73\pm 7\,{\rm hr^{-1}}$), despite LOFAR nominally having substantially better sensitivity.
The RFI environment has the potential to adversely impact single-pulse statistics, and could be a reason why the LOFAR pulse rate is smaller than expected from the MWA measurements. 
Nominally, the MWA is substantially less affected by RFI than LOFAR in certain frequency bands, particularly around 150\,MHz \citep[e.g.][]{2013A&AOffringa,2015PASAOffringa}. 

A caveat to this discussion is that comparing pulse rates between different observing epochs intrinsically assumes that the pulses are produced by a Poisson process, where pulses occur independently and at a constant average rate.
It is unclear whether this is the case for RRAT emission in general.
\citet{2018ApJShapiro-Albert} find that there is evidence for pulse clustering on relatively short time scales (tens of rotations) beyond what can be attributed to a random emission process, whereas \citet{2011MNRASPalliyaguru} did not find such an effect occurring on longer time scales. 
From our analysis, we do not find evidence that single pulse emission from RRAT \rrat\ is anything other than randomly distributed.

\subsection{Pulse peak misalignment}\label{sec:disc_align}
In many instances when comparing our coincident pulses, there is an offset between the peak emission locations.
For example, in Figure~\ref{fig:spgal}, we see that the peaks at 154\,MHz and 1.4\,GHz align reasonably well for pulse 1292 (middle), but do not in pulses 527 (left) and 4318 (right).
Initially, one might assume a clock offset between the two telescopes that has not been taken into account, although this would nominally appear as a constant offset between the peaks of coincident pulses.
Upon careful inspection of matched single pulses, it is interesting to note that while the peaks do not align, there is generally complete overlap in the total emission envelope at each frequency (which is also true for pulses 527 and 4318).
This points towards a phase-dependent spectral index such as those seen for some millisecond pulsars \citep[e.g.][]{2015MNRASDai}.
Another possibility, assuming that the emission frequency is proportional to the emission location (e.g. altitude), is that there are multiple discrete emission zones firing at slightly different times.
Given that characterising the pulse-to-pulse variability of even well-studied pulsars is challenging \citep[e.g. jitter noise,][]{2012MNRASLiu,2014MNRASShannon}, it is difficult to confidently determine why single pulses from a RRAT would exhibit emission peak phase offsets.
Determining the geometry of the system through additional wideband single-pulse polarisation measurements and analysis \citep[e.g.][]{2019MNRASCaleb} would provide stronger evidence as to the nature of the single-pulse misalignment and why it changes from pulse to pulse.

\section{CONCLUDING REMARKS}\label{sec:conclusions}
We have presented the first detection of RRAT \rrat\ made simultaneously over a wide range in frequency (154--1400\,MHz) with the MWA and Parkes radio telescopes.
Generally, an RFI-quiet environment, such as that at the MRO, is much more conducive to single-pulse based statistical work such as what we have presented here.
It is important to make use of multi-telescope, multi-frequency studies of these objects in order to place their emission properties in the context of the normal pulsar population, given that many emission characteristics of RRATs remain a mystery.

Over the respective observations we detected \nMWA\ and \nPKS\ pulses with the MWA and Parkes, implying pulse rates of $73\pm 7$ and $43\pm 5$ pulses per hour.
The single-pulse spectral index distribution of RRAT \rrat\, with a mean of $\alpha_{154}^{1369}=\specidx$ is relatively steep compared to the normal pulsar population, though scintillation bias makes it difficult to robustly estimate from a single observation.
The pulse energy distribution of RRAT \rrat\ is best described by either a log-normal or truncated exponential model, in general agreement with previous RRAT studies, and with the typical pulsar population.
We also provide the first polarimetric profiles and RM estimate for this pulsar, with $\rm RM_{ISM}=3.8\pm 0.1\,rad\,m^{-2}$, where the ionospheric correction is the dominant source of uncertainty for the low-frequency MWA measurement.

This is the first time scintillation properties have been measured for a RRAT, using a necessarily modified version of the standard autocorrelation analysis employed.
Even with the inherently irregular single-pulse emission, we clearly see scintillation at 1.4\,GHz, with a characteristic bandwidth of $\nud=\nudPKS$\,MHz and time scale of $\td=\tdPKS$\,s.
Notwithstanding the limitation of the number of scintles observed at 1.4\,GHz, and the inadequate frequency resolution at 154\,MHz, we place constraints on the scintillation velocity and turbulence within the ISM along the line-of-sight to RRAT \rrat. 
We also measure a scintillation frequency scaling index of $\gamma=-4.2$, which is close to the theoretically steepest value, $\gamma=-4.4$, of Kolmogorov turbulence.
The line-of-sight for this RRAT is relatively typical of nearby pulsars based on the estimated scaling index, and on the turbulence strength ($\cn \lesssim 2.8\times 10^{-4}\,{\rm m^{-20/3}}$), as might be expected given its large Galactic latitude ($b=-60.2^\circ$).
In future scintillation analysis of RRAT \rrat, it will be important to study the emission at an intermediate frequency band, nominally in the 300--700\,MHz range, where the scintillation properties are expected to be better suited to characterisation.

The steep spectrum of the emission and the relatively large pulse rates indicate that future and on-going surveys, such as those using LOFAR, the MWA and those planned with the SKA, will be in a good position to find new examples of these objects.

\begin{acknowledgements}
\noindent{\em People and funding:} The authors would like to thank the anonymous referee for their careful feedback and suggestions.
BWM thanks C. Karako-Argamann and P. Chawla for valuable discussion regarding the pulse rates of \rrat\ and the comparison thereof.
The authors acknowledge the contribution of an Australian Government Research Training Program Scholarship in supporting this research.
RMS also acknowledges support through ARC grant CE170100004.
Part of this research was supported by the Australian Research Council Centre of Excellence for All-sky Astrophysics (CAASTRO), through project number CE110001020.

\noindent{\em Software:} This work made use of the following software packages: PSRCHIVE \citep{2004PASAHotan,2012AR&TvanStraten}, DSPSR \citep{2011PASAvanStraten} and STILTS \citep{2006ASPCTaylor}.
The following Python modules we also used as part of this work: AstroPy \citep{2013A&AAstropy,2018AJAstropy}, NumPy \citep{2011CSEvanderWalt_numpy}, SciPy \citep{2001Jones_scipy}, LMfit \citep{2014Newville_lmfit}, and Matplotlib \citep{2007CSEHunter_matplotlib}.

\noindent{\em Facilities:} This scientific work makes use of the Murchison Radio-astronomy Observatory, operated by CSIRO. 
We acknowledge the Wajarri Yamatji people as the traditional owners of the Observatory site. 
Support for the operation of the MWA is provided by the Australian Government (NCRIS), under a contract to Curtin University administered by Astronomy Australia Limited. 
The Parkes radio telescope is part of the Australia Telescope National Facility, which is funded by the Commonwealth of Australia for operation as a National Facility managed by CSIRO.
We acknowledge the Pawsey Supercomputing Centre which is supported by the Western Australian and Australian Governments. 
\end{acknowledgements}

\bibliographystyle{pasa-mnras}
\bibliography{references}

\begin{thebibliography}{}
\makeatletter
\relax
\def\mn@urlcharsother{\let\do\@makeother \do\$\do\&\do\#\do\^\do\_\do\%\do\~}
\definecolor{darkblue}{rgb}{0,0,0.597656}
\def\mndoi{\begingroup\mn@urlcharsother \@ifnextchar [ {\mndoi@} {\mndoi@[]}}
\def\mndoi@[#1]#2{\def\@tempa{#1}\ifx\@tempa\@empty \href
  {http://dx.doi.org/#2} {\textcolor{darkblue}{doi:#2}}\else \href
  {http://dx.doi.org/#2} {\textcolor{darkblue}{#1}}\fi \endgroup}
\def\mn@eprint#1#2{\mn@eprint@#1:#2::\@nil}
\def\mn@eprint@arXiv#1{\href {http://arxiv.org/abs/#1} {{\tt arXiv:#1}}}
\def\mn@eprint@dblp#1{\href {http://dblp.uni-trier.de/rec/bibtex/#1.xml}
  {dblp:#1}}
\def\mn@eprint@#1:#2:#3:#4\@nil{\def\@tempa {#1}\def\@tempb {#2}\def\@tempc
  {#3}\ifx \@tempc \@empty \let \@tempc \@tempb \let \@tempb \@tempa \fi \ifx
  \@tempb \@empty \def\@tempb {arXiv}\fi \@ifundefined
  {mn@eprint@\@tempb}{\@tempb:\@tempc}{\expandafter \expandafter \csname
  mn@eprint@\@tempb\endcsname \expandafter{\@tempc}}}

\bibitem[\protect\citeauthoryear{{Bates}, {Lorimer}  \& {Verbiest}}{{Bates}
  et~al.}{2013}]{2013MNRASBates}
{Bates} S.~D.,  {Lorimer} D.~R.,   {Verbiest} J.~P.~W.,  2013, \mndoi [\mnras]
  {10.1093/mnras/stt257}, \href
  {https://ui.adsabs.harvard.edu/#abs/2013MNRAS.431.1352B} {431, 1352}

\bibitem[\protect\citeauthoryear{{Bhat}, {Rao}  \& {Gupta}}{{Bhat}
  et~al.}{1999}]{1999ApJSBhat}
{Bhat} N.~D.~R.,  {Rao} A.~P.,   {Gupta} Y.,  1999, \mndoi [\apjs]
  {10.1086/313198}, \href {http://cdsads.u-strasbg.fr/abs/1999ApJS..121..483B}
  {121, 483}

\bibitem[\protect\citeauthoryear{{Bhat}, {Cordes}, {Camilo}, {Nice}  \&
  {Lorimer}}{{Bhat} et~al.}{2004}]{2004ApJBhat}
{Bhat} N.~D.~R.,  {Cordes} J.~M.,  {Camilo} F.,  {Nice} D.~J.,   {Lorimer}
  D.~R.,  2004, \mndoi [\apj] {10.1086/382680}, \href
  {https://ui.adsabs.harvard.edu/#abs/2004ApJ...605..759B} {605, 759}

\bibitem[\protect\citeauthoryear{{Bhat}, {Tingay}  \& {Knight}}{{Bhat}
  et~al.}{2008}]{2008ApJBhat}
{Bhat} N.~D.~R.,  {Tingay} S.~J.,   {Knight} H.~S.,  2008, \mndoi [\apj]
  {10.1086/528735}, \href
  {https://ui.adsabs.harvard.edu/#abs/2008ApJ...676.1200B} {676, 1200}

\bibitem[\protect\citeauthoryear{{Bhat}, {Ord}, {Tremblay}, {McSweeney}  \&
  {Tingay}}{{Bhat} et~al.}{2016}]{2016ApJBhat}
{Bhat} N.~D.~R.,  {Ord} S.~M.,  {Tremblay} S.~E.,  {McSweeney} S.~J.,
  {Tingay} S.~J.,  2016, \mndoi [\apj] {10.3847/0004-637X/818/1/86}, \href
  {http://adsabs.harvard.edu/abs/2016ApJ...818...86B} {818, 86}

\bibitem[\protect\citeauthoryear{{Bhat} et~al.,}{{Bhat}
  et~al.}{2018}]{2018ApJSBhat}
{Bhat} N.~D.~R.,  et~al., 2018, \mndoi [\apjs] {10.3847/1538-4365/aad37c},
  \href {https://ui.adsabs.harvard.edu/#abs/2018ApJS..238....1B} {238, 1}

\bibitem[\protect\citeauthoryear{{Bhattacharyya} et~al.,}{{Bhattacharyya}
  et~al.}{2018}]{2018MNRASBhattacharyya}
{Bhattacharyya} B.,  et~al., 2018, \mndoi [\mnras] {10.1093/mnras/sty923},
  \href {https://ui.adsabs.harvard.edu/#abs/2018MNRAS.477.4090B} {477, 4090}

\bibitem[\protect\citeauthoryear{{Boyles} et~al.,}{{Boyles}
  et~al.}{2013}]{2013ApJBoyles}
{Boyles} J.,  et~al., 2013, \mndoi [\apj] {10.1088/0004-637X/763/2/80}, \href
  {https://ui.adsabs.harvard.edu/#abs/2013ApJ...763...80B} {763, 80}

\bibitem[\protect\citeauthoryear{{Brentjens} \& {de Bruyn}}{{Brentjens} \& {de
  Bruyn}}{2005}]{2005A&ABrentjens}
{Brentjens} M.~A.,  {de Bruyn} A.~G.,  2005, \mndoi [\aap]
  {10.1051/0004-6361:20052990}, \href
  {http://adsabs.harvard.edu/abs/2005A%26A...441.1217B} {441, 1217}

\bibitem[\protect\citeauthoryear{{Burke-Spolaor} \& {Bailes}}{{Burke-Spolaor}
  \& {Bailes}}{2010}]{2010MNRASBurke-Spolaor}
{Burke-Spolaor} S.,  {Bailes} M.,  2010, \mndoi [\mnras]
  {10.1111/j.1365-2966.2009.15965.x}, \href
  {https://ui.adsabs.harvard.edu/#abs/2010MNRAS.402..855B} {402, 855}

\bibitem[\protect\citeauthoryear{{Burke-Spolaor} et~al.,}{{Burke-Spolaor}
  et~al.}{2012}]{2012MNRASBurke-Spolaor}
{Burke-Spolaor} S.,  et~al., 2012, \mndoi [\mnras]
  {10.1111/j.1365-2966.2012.20998.x}, \href
  {http://adsabs.harvard.edu/abs/2012MNRAS.423.1351B} {423, 1351}

\bibitem[\protect\citeauthoryear{{Caleb} et~al.,}{{Caleb}
  et~al.}{2019}]{2019MNRASCaleb}
{Caleb} M.,  et~al., 2019, \mndoi [\mnras] {10.1093/mnras/stz1352}, \href
  {https://ui.adsabs.harvard.edu/abs/2019MNRAS.tmp.1287C} {p.~1287}

\bibitem[\protect\citeauthoryear{{Cordes}}{{Cordes}}{1986}]{1986ApJCordes}
{Cordes} J.~M.,  1986, \mndoi [\apj] {10.1086/164764}, \href
  {https://ui.adsabs.harvard.edu/\#abs/1986ApJ...311..183C} {311, 183}

\bibitem[\protect\citeauthoryear{{Cordes} \& {Lazio}}{{Cordes} \&
  {Lazio}}{2002}]{2002ArXivCordes}
{Cordes} J.~M.,  {Lazio} T.~J.~W.,  2002, preprint, \href
  {https://ui.adsabs.harvard.edu/#abs/2002astro.ph..7156C} {} (\mn@eprint
  {arXiv} {astro-ph/0207156})

\bibitem[\protect\citeauthoryear{{Cordes} \& {Lazio}}{{Cordes} \&
  {Lazio}}{2003}]{2003ArXivCordes}
{Cordes} J.~M.,  {Lazio} T.~J.~W.,  2003, preprint, \href
  {https://ui.adsabs.harvard.edu/#abs/2003astro.ph..1598C} {} (\mn@eprint
  {arXiv} {astro-ph/0301598})

\bibitem[\protect\citeauthoryear{{Cordes} \& {Rickett}}{{Cordes} \&
  {Rickett}}{1998}]{1998ApJCordes}
{Cordes} J.~M.,  {Rickett} B.~J.,  1998, \mndoi [\apj] {10.1086/306358}, \href
  {https://ui.adsabs.harvard.edu/#abs/1998ApJ...507..846C} {507, 846}

\bibitem[\protect\citeauthoryear{{Cordes} \& {Shannon}}{{Cordes} \&
  {Shannon}}{2008}]{2008ApJCordes}
{Cordes} J.~M.,  {Shannon} R.~M.,  2008, \mndoi [\apj] {10.1086/589425}, \href
  {https://ui.adsabs.harvard.edu/#abs/2008ApJ...682.1152C} {682, 1152}

\bibitem[\protect\citeauthoryear{{Cordes}, {Wolszczan}, {Dewey}, {Blaskiewicz}
  \& {Stinebring}}{{Cordes} et~al.}{1990}]{1990ApJCordes}
{Cordes} J.~M.,  {Wolszczan} A.,  {Dewey} R.~J.,  {Blaskiewicz} M.,
  {Stinebring} D.~R.,  1990, \mndoi [\apj] {10.1086/168310}, \href
  {https://ui.adsabs.harvard.edu/#abs/1990ApJ...349..245C} {349, 245}

\bibitem[\protect\citeauthoryear{{Cordes}, {Bhat}, {Hankins}, {McLaughlin}  \&
  {Kern}}{{Cordes} et~al.}{2004}]{2004ApJCordes}
{Cordes} J.~M.,  {Bhat} N.~D.~R.,  {Hankins} T.~H.,  {McLaughlin} M.~A.,
  {Kern} J.,  2004, \mndoi [\apj] {10.1086/422495}, \href
  {https://ui.adsabs.harvard.edu/#abs/2004ApJ...612..375C} {612, 375}

\bibitem[\protect\citeauthoryear{{Cui}, {Boyles}, {McLaughlin}  \&
  {Palliyaguru}}{{Cui} et~al.}{2017}]{2017ApJCui}
{Cui} B.~Y.,  {Boyles} J.,  {McLaughlin} M.~A.,   {Palliyaguru} N.,  2017,
  \mndoi [\apj] {10.3847/1538-4357/aa6aa9}, \href
  {https://ui.adsabs.harvard.edu/#abs/2017ApJ...840....5C} {840, 5}

\bibitem[\protect\citeauthoryear{{Dai} et~al.,}{{Dai}
  et~al.}{2015}]{2015MNRASDai}
{Dai} S.,  et~al., 2015, \mndoi [\mnras] {10.1093/mnras/stv508}, \href
  {https://ui.adsabs.harvard.edu/abs/2015MNRAS.449.3223D} {449, 3223}

\bibitem[\protect\citeauthoryear{{Deller}, {Weisberg}, {Nice}  \&
  {Chatterjee}}{{Deller} et~al.}{2018}]{2018ApJDeller}
{Deller} A.~T.,  {Weisberg} J.~M.,  {Nice} D.~J.,   {Chatterjee} S.,  2018,
  \mndoi [\apj] {10.3847/1538-4357/aacf95}, \href
  {https://ui.adsabs.harvard.edu/\#abs/2018ApJ...862..139D} {862, 139}

\bibitem[\protect\citeauthoryear{{Everett} \& {Weisberg}}{{Everett} \&
  {Weisberg}}{2001}]{2001ApJEverett}
{Everett} J.~E.,  {Weisberg} J.~M.,  2001, \mndoi [\apj] {10.1086/320652},
  \href {http://adsabs.harvard.edu/abs/2001ApJ...553..341E} {553, 341}

\bibitem[\protect\citeauthoryear{{Gajjar}, {Joshi}  \& {Kramer}}{{Gajjar}
  et~al.}{2012}]{2012MNRASGajjar}
{Gajjar} V.,  {Joshi} B.~C.,   {Kramer} M.,  2012, \mndoi [\mnras]
  {10.1111/j.1365-2966.2012.21296.x}, \href
  {https://ui.adsabs.harvard.edu/\#abs/2012MNRAS.424.1197G} {424, 1197}

\bibitem[\protect\citeauthoryear{{Gonzalez} et~al.,}{{Gonzalez}
  et~al.}{2011}]{2011ApJGonzalez}
{Gonzalez} M.~E.,  et~al., 2011, \mndoi [\apj] {10.1088/0004-637X/743/2/102},
  \href {https://ui.adsabs.harvard.edu/\#abs/2011ApJ...743..102G} {743, 102}

\bibitem[\protect\citeauthoryear{{Gould} \& {Lyne}}{{Gould} \&
  {Lyne}}{1998}]{1998MNRASGould}
{Gould} D.~M.,  {Lyne} A.~G.,  1998, \mndoi [\mnras]
  {10.1046/j.1365-8711.1998.02018.x}, \href
  {https://ui.adsabs.harvard.edu/\#abs/1998MNRAS.301..235G} {301, 235}

\bibitem[\protect\citeauthoryear{{Gupta}, {Rickett}  \& {Lyne}}{{Gupta}
  et~al.}{1994}]{1994MNRASGupta}
{Gupta} Y.,  {Rickett} B.~J.,   {Lyne} A.~G.,  1994, \mndoi [\mnras]
  {10.1093/mnras/269.4.1035}, \href
  {https://ui.adsabs.harvard.edu/#abs/1994MNRAS.269.1035G} {269, 1035}

\bibitem[\protect\citeauthoryear{{Han}, {Manchester}, {van Straten}  \&
  {Demorest}}{{Han} et~al.}{2018}]{2018ApJSHan}
{Han} J.~L.,  {Manchester} R.~N.,  {van Straten} W.,   {Demorest} P.,  2018,
  \mndoi [\apjs] {10.3847/1538-4365/aa9c45}, \href
  {https://ui.adsabs.harvard.edu/#abs/2018ApJS..234...11H} {234, 11}

\bibitem[\protect\citeauthoryear{{Hern{\'a}ndez-Pajares}
  et~al.,}{{Hern{\'a}ndez-Pajares} et~al.}{2009}]{2009JGeodHernandez-Pajares}
{Hern{\'a}ndez-Pajares} M.,  et~al., 2009, \mndoi [Journal of Geodesy]
  {10.1007/s00190-008-0266-1}, \href
  {https://ui.adsabs.harvard.edu/#abs/2009JGeod..83..263H} {83, 263}

\bibitem[\protect\citeauthoryear{{Hotan}, {van Straten}  \&
  {Manchester}}{{Hotan} et~al.}{2004}]{2004PASAHotan}
{Hotan} A.~W.,  {van Straten} W.,   {Manchester} R.~N.,  2004, \mndoi [\pasa]
  {10.1071/AS04022}, \href {http://adsabs.harvard.edu/abs/2004PASA...21..302H}
  {21, 302}

\bibitem[\protect\citeauthoryear{Hunter}{Hunter}{2007}]{2007CSEHunter_matplotlib}
Hunter J.~D.,  2007, \mndoi [Computing In Science \& Engineering]
  {10.1109/MCSE.2007.55}, 9, 90

\bibitem[\protect\citeauthoryear{{Jankowski}, {van Straten}, {Keane}, {Bailes},
  {Barr}, {Johnston}  \& {Kerr}}{{Jankowski} et~al.}{2018}]{2018MNRASJankowski}
{Jankowski} F.,  {van Straten} W.,  {Keane} E.~F.,  {Bailes} M.,  {Barr} E.~D.,
   {Johnston} S.,   {Kerr} M.,  2018, \mndoi [\mnras] {10.1093/mnras/stx2476},
  \href {http://adsabs.harvard.edu/abs/2018MNRAS.473.4436J} {473, 4436}

\bibitem[\protect\citeauthoryear{{Janssen}, {Stappers}, {Bassa}, {Cognard},
  {Kramer}  \& {Theureau}}{{Janssen} et~al.}{2010}]{2010A&AJanssen}
{Janssen} G.~H.,  {Stappers} B.~W.,  {Bassa} C.~G.,  {Cognard} I.,  {Kramer}
  M.,   {Theureau} G.,  2010, \mndoi [\aap] {10.1051/0004-6361/200911728},
  \href {https://ui.adsabs.harvard.edu/\#abs/2010A&A...514A..74J} {514, A74}

\bibitem[\protect\citeauthoryear{{Jennings}, {Kaplan}, {Chatterjee}, {Cordes}
  \& {Deller}}{{Jennings} et~al.}{2018}]{2018ApJJennings}
{Jennings} R.~J.,  {Kaplan} D.~L.,  {Chatterjee} S.,  {Cordes} J.~M.,
  {Deller} A.~T.,  2018, \mndoi [\apj] {10.3847/1538-4357/aad084}, \href
  {https://ui.adsabs.harvard.edu/\#abs/2018ApJ...864...26J} {864, 26}

\bibitem[\protect\citeauthoryear{{Johnston} \& {Kerr}}{{Johnston} \&
  {Kerr}}{2018}]{2018MNRAJohnston}
{Johnston} S.,  {Kerr} M.,  2018, \mndoi [\mnras] {10.1093/mnras/stx3095},
  \href {https://ui.adsabs.harvard.edu/\#abs/2018MNRAS.474.4629J} {474, 4629}

\bibitem[\protect\citeauthoryear{{Johnston}, {Nicastro}  \&
  {Koribalski}}{{Johnston} et~al.}{1998}]{1998MNRASJohnston}
{Johnston} S.,  {Nicastro} L.,   {Koribalski} B.,  1998, \mndoi [\mnras]
  {10.1046/j.1365-8711.1998.01461.x}, \href
  {http://adsabs.harvard.edu/abs/1998MNRAS.297..108J} {297, 108}

\bibitem[\protect\citeauthoryear{Jones, Oliphant, Peterson  \& Others}{Jones
  et~al.}{2001}]{2001Jones_scipy}
Jones E.,  Oliphant T.,  Peterson P.,   Others 2001, {SciPy: Open source
  scientific tools for Python}, \url {https://www.scipy.org/}

\bibitem[\protect\citeauthoryear{{Karako-Argaman} et~al.,}{{Karako-Argaman}
  et~al.}{2015}]{2015ApJKarako-Argaman}
{Karako-Argaman} C.,  et~al., 2015, \mndoi [\apj] {10.1088/0004-637X/809/1/67},
  \href {https://ui.adsabs.harvard.edu/#abs/2015ApJ...809...67K} {809, 67}

\bibitem[\protect\citeauthoryear{{Karastergiou}, {Hotan}, {van Straten},
  {McLaughlin}  \& {Ord}}{{Karastergiou} et~al.}{2009}]{2009MNRASKarastergiou}
{Karastergiou} A.,  {Hotan} A.~W.,  {van Straten} W.,  {McLaughlin} M.~A.,
  {Ord} S.~M.,  2009, \mndoi [\mnras] {10.1111/j.1745-3933.2009.00671.x}, \href
  {https://ui.adsabs.harvard.edu/\#abs/2009MNRAS.396L..95K} {396, L95}

\bibitem[\protect\citeauthoryear{{Karuppusamy}, {Stappers}  \& {van
  Straten}}{{Karuppusamy} et~al.}{2010}]{2010A&AKaruppusamy}
{Karuppusamy} R.,  {Stappers} B.~W.,   {van Straten} W.,  2010, \mndoi [\aap]
  {10.1051/0004-6361/200913729}, \href
  {https://ui.adsabs.harvard.edu/#abs/2010A&A...515A..36K} {515, A36}

\bibitem[\protect\citeauthoryear{{Keane}}{{Keane}}{2016}]{2016MNRASKeane}
{Keane} E.~F.,  2016, \mndoi [\mnras] {10.1093/mnras/stw767}, \href
  {https://ui.adsabs.harvard.edu/#abs/2016MNRAS.459.1360K} {459, 1360}

\bibitem[\protect\citeauthoryear{{Keane}, {Kramer}, {Lyne}, {Stappers}  \&
  {McLaughlin}}{{Keane} et~al.}{2011}]{2011MNRASKeane}
{Keane} E.~F.,  {Kramer} M.,  {Lyne} A.~G.,  {Stappers} B.~W.,   {McLaughlin}
  M.~A.,  2011, \mndoi [\mnras] {10.1111/j.1365-2966.2011.18917.x}, \href
  {https://ui.adsabs.harvard.edu/#abs/2011MNRAS.415.3065K} {415, 3065}

\bibitem[\protect\citeauthoryear{{Kramer}, {Karastergiou}, {Gupta}, {Johnston},
  {Bhat}  \& {Lyne}}{{Kramer} et~al.}{2003}]{2003A&AKramer}
{Kramer} M.,  {Karastergiou} A.,  {Gupta} Y.,  {Johnston} S.,  {Bhat} N.~D.~R.,
    {Lyne} A.~G.,  2003, \mndoi [\aap] {10.1051/0004-6361:20030842}, \href
  {https://ui.adsabs.harvard.edu/#abs/2003A&A...407..655K} {407, 655}

\bibitem[\protect\citeauthoryear{{Levin} et~al.,}{{Levin}
  et~al.}{2016}]{2016ApJLevin}
{Levin} L.,  et~al., 2016, \mndoi [\apj] {10.3847/0004-637X/818/2/166}, \href
  {https://ui.adsabs.harvard.edu/\#abs/2016ApJ...818..166L} {818, 166}

\bibitem[\protect\citeauthoryear{{Li}}{{Li}}{2006}]{2006ApJLi}
{Li} X.-D.,  2006, \mndoi [\apj] {10.1086/506962}, \href
  {https://ui.adsabs.harvard.edu/#abs/2006ApJ...646L.139L} {646, L139}

\bibitem[\protect\citeauthoryear{{Li}, {Spitkovsky}  \& {Tchekhovskoy}}{{Li}
  et~al.}{2012}]{2012ApJLi}
{Li} J.,  {Spitkovsky} A.,   {Tchekhovskoy} A.,  2012, \mndoi [\apj]
  {10.1088/2041-8205/746/2/L24}, \href
  {https://ui.adsabs.harvard.edu/#abs/2012ApJ...746L..24L} {746, L24}

\bibitem[\protect\citeauthoryear{{Liu}, {Keane}, {Lee}, {Kramer}, {Cordes}  \&
  {Purver}}{{Liu} et~al.}{2012}]{2012MNRASLiu}
{Liu} K.,  {Keane} E.~F.,  {Lee} K.~J.,  {Kramer} M.,  {Cordes} J.~M.,
  {Purver} M.~B.,  2012, \mndoi [\mnras] {10.1111/j.1365-2966.2011.20041.x},
  \href {https://ui.adsabs.harvard.edu/\#abs/2012MNRAS.420..361L} {420, 361}

\bibitem[\protect\citeauthoryear{{Lynch} et~al.,}{{Lynch}
  et~al.}{2013}]{2013ApJLynch}
{Lynch} R.~S.,  et~al., 2013, \mndoi [\apj] {10.1088/0004-637X/763/2/81}, \href
  {https://ui.adsabs.harvard.edu/#abs/2013ApJ...763...81L} {763, 81}

\bibitem[\protect\citeauthoryear{{Lyne} \& {Smith}}{{Lyne} \&
  {Smith}}{1982}]{1982NaturLyne}
{Lyne} A.~G.,  {Smith} F.~G.,  1982, \mndoi [\nat] {10.1038/298825a0}, \href
  {https://ui.adsabs.harvard.edu/\#abs/1982Natur.298..825L} {298, 825}

\bibitem[\protect\citeauthoryear{{Manchester}, {Han}  \& {Qiao}}{{Manchester}
  et~al.}{1998}]{1998MNRASManchester}
{Manchester} R.~N.,  {Han} J.~L.,   {Qiao} G.~J.,  1998, \mndoi [\mnras]
  {10.1046/j.1365-8711.1998.01204.x}, \href
  {https://ui.adsabs.harvard.edu/\#abs/1998MNRAS.295..280M} {295, 280}

\bibitem[\protect\citeauthoryear{{Manchester} et~al.,}{{Manchester}
  et~al.}{2013}]{2013PASAManchester}
{Manchester} R.~N.,  et~al., 2013, \mndoi [Publications of the Astronomical
  Society of Australia] {10.1017/pasa.2012.017}, \href
  {https://ui.adsabs.harvard.edu/#abs/2013PASA...30...17M} {30, e017}

\bibitem[\protect\citeauthoryear{{Maron}, {Kijak}, {Kramer}  \&
  {Wielebinski}}{{Maron} et~al.}{2000}]{2000A&ASMaron}
{Maron} O.,  {Kijak} J.,  {Kramer} M.,   {Wielebinski} R.,  2000, \mndoi
  [\aaps] {10.1051/aas:2000298}, \href
  {https://ui.adsabs.harvard.edu/#abs/2000A&AS..147..195M} {147, 195}

\bibitem[\protect\citeauthoryear{{McLaughlin} et~al.,}{{McLaughlin}
  et~al.}{2006}]{2006NatureMcLaughlin}
{McLaughlin} M.~A.,  et~al., 2006, \mndoi [\nat] {10.1038/nature04440}, \href
  {http://adsabs.harvard.edu/abs/2006Natur.439..817M} {439, 817}

\bibitem[\protect\citeauthoryear{{McLaughlin} et~al.,}{{McLaughlin}
  et~al.}{2009}]{2009MNRASMcLaughlin}
{McLaughlin} M.~A.,  et~al., 2009, \mndoi [\mnras]
  {10.1111/j.1365-2966.2009.15584.x}, \href
  {https://ui.adsabs.harvard.edu/#abs/2009MNRAS.400.1431M} {400, 1431}

\bibitem[\protect\citeauthoryear{{Melrose} \& {Yuen}}{{Melrose} \&
  {Yuen}}{2014}]{2014MNRASMelrose}
{Melrose} D.~B.,  {Yuen} R.,  2014, \mndoi [\mnras] {10.1093/mnras/stt1876},
  \href {https://ui.adsabs.harvard.edu/#abs/2014MNRAS.437..262M} {437, 262}

\bibitem[\protect\citeauthoryear{{Meyers} et~al.,}{{Meyers}
  et~al.}{2017}]{2017ApJMeyers}
{Meyers} B.~W.,  et~al., 2017, \mndoi [\apj] {10.3847/1538-4357/aa8bba}, \href
  {http://adsabs.harvard.edu/abs/2017ApJ...851...20M} {851, 20}

\bibitem[\protect\citeauthoryear{{Meyers} et~al.,}{{Meyers}
  et~al.}{2018}]{2018ApJMeyers}
{Meyers} B.~W.,  et~al., 2018, \mndoi [\apj] {10.3847/1538-4357/aaee7b}, \href
  {https://ui.adsabs.harvard.edu/\#abs/2018ApJ...869..134M} {869, 134}

\bibitem[\protect\citeauthoryear{{Michel} \& {Dessler}}{{Michel} \&
  {Dessler}}{1981}]{1981ApJMichel}
{Michel} F.~C.,  {Dessler} A.~J.,  1981, \mndoi [\apj] {10.1086/159511}, \href
  {https://ui.adsabs.harvard.edu/#abs/1981ApJ...251..654M} {251, 654}

\bibitem[\protect\citeauthoryear{{Mickaliger} et~al.,}{{Mickaliger}
  et~al.}{2012}]{2012ApJMickaliger}
{Mickaliger} M.~B.,  et~al., 2012, \mndoi [\apj] {10.1088/0004-637X/760/1/64},
  \href {https://ui.adsabs.harvard.edu/#abs/2012ApJ...760...64M} {760, 64}

\bibitem[\protect\citeauthoryear{{Mickaliger}, {McEwen}, {McLaughlin}  \&
  {Lorimer}}{{Mickaliger} et~al.}{2018}]{2018MNRASMickaliger}
{Mickaliger} M.~B.,  {McEwen} A.~E.,  {McLaughlin} M.~A.,   {Lorimer} D.~R.,
  2018, \mndoi [\mnras] {10.1093/mnras/sty1785}, \href
  {https://ui.adsabs.harvard.edu/#abs/2018MNRAS.479.5413M} {479, 5413}

\bibitem[\protect\citeauthoryear{{Mitchell}, {Greenhill}, {Wayth}, {Sault},
  {Lonsdale}, {Cappallo}, {Morales}  \& {Ord}}{{Mitchell}
  et~al.}{2008}]{2008ISTSPMitchell}
{Mitchell} D.~A.,  {Greenhill} L.~J.,  {Wayth} R.~B.,  {Sault} R.~J.,
  {Lonsdale} C.~J.,  {Cappallo} R.~J.,  {Morales} M.~F.,   {Ord} S.~M.,  2008,
  \mndoi [IEEE Journal of Selected Topics in Signal Processing]
  {10.1109/JSTSP.2008.2005327}, \href
  {http://adsabs.harvard.edu/abs/2008ISTSP...2..707M} {2, 707}

\bibitem[\protect\citeauthoryear{{Mitra}, {Wielebinski}, {Kramer}  \&
  {Jessner}}{{Mitra} et~al.}{2003}]{2003A&AMitra}
{Mitra} D.,  {Wielebinski} R.,  {Kramer} M.,   {Jessner} A.,  2003, \mndoi
  [\aap] {10.1051/0004-6361:20021702}, \href
  {https://ui.adsabs.harvard.edu/\#abs/2003A&A...398..993M} {398, 993}

\bibitem[\protect\citeauthoryear{{Mitra}, {Basu}, {Maciesiak}, {Skrzypczak},
  {Melikidze}, {Szary}  \& {Krzeszowski}}{{Mitra} et~al.}{2016}]{2016ApJMitra}
{Mitra} D.,  {Basu} R.,  {Maciesiak} K.,  {Skrzypczak} A.,  {Melikidze} G.~I.,
  {Szary} A.,   {Krzeszowski} K.,  2016, \mndoi [\apj]
  {10.3847/1538-4357/833/1/28}, \href
  {https://ui.adsabs.harvard.edu/\#abs/2016ApJ...833...28M} {833, 28}

\bibitem[\protect\citeauthoryear{Newville, Stensitzki, Allen  \&
  Ingargiola}{Newville et~al.}{2014}]{2014Newville_lmfit}
Newville M.,  Stensitzki T.,  Allen D.~B.,   Ingargiola A.,  2014, {LMFIT:
  Non-Linear Least-Square Minimization and Curve-Fitting for Python},
  \mndoi{10.5281/zenodo.11813}, \url {https://doi.org/10.5281/zenodo.11813}

\bibitem[\protect\citeauthoryear{{Noutsos}, {Johnston}, {Kramer}  \&
  {Karastergiou}}{{Noutsos} et~al.}{2008}]{2008MNRASNoutsos}
{Noutsos} A.,  {Johnston} S.,  {Kramer} M.,   {Karastergiou} A.,  2008, \mndoi
  [\mnras] {10.1111/j.1365-2966.2008.13188.x}, \href
  {http://adsabs.harvard.edu/abs/2008MNRAS.386.1881N} {386, 1881}

\bibitem[\protect\citeauthoryear{{Offringa} et~al.,}{{Offringa}
  et~al.}{2013}]{2013A&AOffringa}
{Offringa} A.~R.,  et~al., 2013, \mndoi [\aap] {10.1051/0004-6361/201220293},
  \href {https://ui.adsabs.harvard.edu/\#abs/2013A&A...549A..11O} {549, A11}

\bibitem[\protect\citeauthoryear{{Offringa} et~al.,}{{Offringa}
  et~al.}{2015}]{2015PASAOffringa}
{Offringa} A.~R.,  et~al., 2015, \mndoi [Publications of the Astronomical
  Society of Australia] {10.1017/pasa.2015.7}, \href
  {https://ui.adsabs.harvard.edu/\#abs/2015PASA...32....8O} {32, e008}

\bibitem[\protect\citeauthoryear{{Ord}, {van Straten}, {Hotan}  \&
  {Bailes}}{{Ord} et~al.}{2004}]{2004MNRASOrd}
{Ord} S.~M.,  {van Straten} W.,  {Hotan} A.~W.,   {Bailes} M.,  2004, \mndoi
  [\mnras] {10.1111/j.1365-2966.2004.07963.x}, \href
  {https://ui.adsabs.harvard.edu/\#abs/2004MNRAS.352..804O} {352, 804}

\bibitem[\protect\citeauthoryear{{Ord} et~al.,}{{Ord}
  et~al.}{2015}]{2015PASAOrd}
{Ord} S.~M.,  et~al., 2015, \mndoi [\pasa] {10.1017/pasa.2015.5}, \href
  {https://ui.adsabs.harvard.edu/#abs/2015PASA...32....6O} {32, e006}

\bibitem[\protect\citeauthoryear{{Ord}, {Tremblay}, {McSweeney}, {Bhat},
  {Sobey}, {Mitchell}, {Hancock}  \& {Kirsten}}{{Ord}
  et~al.}{2019}]{2019arXivOrd}
{Ord} S.~M.,  {Tremblay} S.~E.,  {McSweeney} S.~J.,  {Bhat} N.~D.~R.,  {Sobey}
  C.,  {Mitchell} D.~A.,  {Hancock} P.~J.,   {Kirsten} F.,  2019, arXiv
  e-prints, \href {https://ui.adsabs.harvard.edu/abs/2019arXiv190501826O} {p.
  arXiv:1905.01826}

\bibitem[\protect\citeauthoryear{{Palliyaguru} et~al.,}{{Palliyaguru}
  et~al.}{2011}]{2011MNRASPalliyaguru}
{Palliyaguru} N.~T.,  et~al., 2011, \mndoi [\mnras]
  {10.1111/j.1365-2966.2011.19388.x}, \href
  {https://ui.adsabs.harvard.edu/#abs/2011MNRAS.417.1871P} {417, 1871}

\bibitem[\protect\citeauthoryear{{Phillips} \& {Clegg}}{{Phillips} \&
  {Clegg}}{1992}]{1992NatPhillips}
{Phillips} J.~A.,  {Clegg} A.~W.,  1992, \mndoi [\nat] {10.1038/360137a0},
  \href {https://ui.adsabs.harvard.edu/#abs/1992Natur.360..137P} {360, 137}

\bibitem[\protect\citeauthoryear{{Redman} \& {Rankin}}{{Redman} \&
  {Rankin}}{2009}]{2009MNRASRedman}
{Redman} S.~L.,  {Rankin} J.~M.,  2009, \mndoi [\mnras]
  {10.1111/j.1365-2966.2009.14632.x}, \href
  {https://ui.adsabs.harvard.edu/\#abs/2009MNRAS.395.1529R} {395, 1529}

\bibitem[\protect\citeauthoryear{{Rickett}}{{Rickett}}{1990}]{1990ARA&ARickett}
{Rickett} B.~J.,  1990, \mndoi [Annual Review of Astronomy and Astrophysics]
  {10.1146/annurev.aa.28.090190.003021}, \href
  {https://ui.adsabs.harvard.edu/\#abs/1990ARA&A..28..561R} {28, 561}

\bibitem[\protect\citeauthoryear{{Shannon} et~al.,}{{Shannon}
  et~al.}{2014}]{2014MNRASShannon}
{Shannon} R.~M.,  et~al., 2014, \mndoi [\mnras] {10.1093/mnras/stu1213}, \href
  {https://ui.adsabs.harvard.edu/abs/2014MNRAS.443.1463S} {443, 1463}

\bibitem[\protect\citeauthoryear{{Shapiro-Albert}, {McLaughlin}  \&
  {Keane}}{{Shapiro-Albert} et~al.}{2018}]{2018ApJShapiro-Albert}
{Shapiro-Albert} B.~J.,  {McLaughlin} M.~A.,   {Keane} E.~F.,  2018, \mndoi
  [\apj] {10.3847/1538-4357/aae2b2}, \href
  {https://ui.adsabs.harvard.edu/#abs/2018ApJ...866..152S} {866, 152}

\bibitem[\protect\citeauthoryear{{Sobey} et~al.,}{{Sobey}
  et~al.}{2019}]{2019MNRASSobey}
{Sobey} C.,  et~al., 2019, \mndoi [\mnras] {10.1093/mnras/stz214}, \href
  {https://ui.adsabs.harvard.edu/\#abs/2019MNRAS.484.3646S} {484, 3646}

\bibitem[\protect\citeauthoryear{{Sotomayor-Beltran}
  et~al.,}{{Sotomayor-Beltran} et~al.}{2013}]{2013A&ASotomayor-Beltran}
{Sotomayor-Beltran} C.,  et~al., 2013, \mndoi [\aap]
  {10.1051/0004-6361/201220728}, \href
  {http://adsabs.harvard.edu/abs/2013A%26A...552A..58S} {552, A58}

\bibitem[\protect\citeauthoryear{{Stappers} et~al.,}{{Stappers}
  et~al.}{2011}]{2011A&AStappers}
{Stappers} B.~W.,  et~al., 2011, \mndoi [\aap] {10.1051/0004-6361/201116681},
  \href {https://ui.adsabs.harvard.edu/#abs/2011A&A...530A..80S} {530, A80}

\bibitem[\protect\citeauthoryear{{Stovall} et~al.,}{{Stovall}
  et~al.}{2015}]{2015ApJStovall}
{Stovall} K.,  et~al., 2015, \mndoi [\apj] {10.1088/0004-637X/808/2/156}, \href
  {https://ui.adsabs.harvard.edu/#abs/2015ApJ...808..156S} {808, 156}

\bibitem[\protect\citeauthoryear{{Sutinjo}, {O'Sullivan}, {Lenc}, {Wayth},
  {Padhi}, {Hall}  \& {Tingay}}{{Sutinjo} et~al.}{2015}]{2015RaScSutinjo}
{Sutinjo} A.,  {O'Sullivan} J.,  {Lenc} E.,  {Wayth} R.~B.,  {Padhi} S.,
  {Hall} P.,   {Tingay} S.~J.,  2015, \mndoi [Radio Science]
  {10.1002/2014RS005517}, \href
  {http://adsabs.harvard.edu/abs/2015RaSc...50...52S} {50, 52}

\bibitem[\protect\citeauthoryear{{Taylor}}{{Taylor}}{2006}]{2006ASPCTaylor}
{Taylor} M.~B.,  2006, in {Gabriel} C.,  {Arviset} C.,  {Ponz} D.,   {Enrique}
  S.,  eds,  Astronomical Society of the Pacific Conference Series Vol. 351,
  Astronomical Data Analysis Software and Systems XV. p.~666

\bibitem[\protect\citeauthoryear{{Taylor} et~al.,}{{Taylor}
  et~al.}{2012}]{2012JAITaylor}
{Taylor} G.~B.,  et~al., 2012, \mndoi [Journal of Astronomical Instrumentation]
  {10.1142/S2251171712500043}, \href
  {https://ui.adsabs.harvard.edu/#abs/2012JAI.....150004T} {1, 1250004}

\bibitem[\protect\citeauthoryear{{Taylor}, {Stovall}, {McCrackan},
  {McLaughlin}, {Miller}, {Karako-Argaman}, {Dowell}  \& {Schinzel}}{{Taylor}
  et~al.}{2016}]{2016ApJTaylor}
{Taylor} G.~B.,  {Stovall} K.,  {McCrackan} M.,  {McLaughlin} M.~A.,  {Miller}
  R.,  {Karako-Argaman} C.,  {Dowell} J.,   {Schinzel} F.~K.,  2016, \mndoi
  [\apj] {10.3847/0004-637X/831/2/140}, \href
  {https://ui.adsabs.harvard.edu/#abs/2016ApJ...831..140T} {831, 140}

\bibitem[\protect\citeauthoryear{{The Astropy Collaboration} et~al.,}{{The
  Astropy Collaboration} et~al.}{2013}]{2013A&AAstropy}
{The Astropy Collaboration} et~al., 2013, \mndoi [\aap]
  {10.1051/0004-6361/201322068}, \href
  {http://adsabs.harvard.edu/abs/2013A%26A...558A..33A} {558, A33}

\bibitem[\protect\citeauthoryear{{The Astropy Collaboration} et~al.,}{{The
  Astropy Collaboration} et~al.}{2018}]{2018AJAstropy}
{The Astropy Collaboration} et~al., 2018, \mndoi [\aj]
  {10.3847/1538-3881/aabc4f}, \href
  {http://adsabs.harvard.edu/abs/2018AJ....156..123T} {156, 123}

\bibitem[\protect\citeauthoryear{{Th{\'e}bault} et~al.,}{{Th{\'e}bault}
  et~al.}{2015}]{2015EP&SThebault}
{Th{\'e}bault} E.,  et~al., 2015, \mndoi [Earth, Planets, and Space]
  {10.1186/s40623-015-0228-9}, \href
  {https://ui.adsabs.harvard.edu/#abs/2015EP&S...67...79T} {67, 79}

\bibitem[\protect\citeauthoryear{{Timokhin}}{{Timokhin}}{2010}]{2010MNRASTimokhin}
{Timokhin} A.~N.,  2010, \mndoi [\mnras] {10.1111/j.1745-3933.2010.00924.x},
  \href {https://ui.adsabs.harvard.edu/#abs/2010MNRAS.408L..41T} {408, L41}

\bibitem[\protect\citeauthoryear{{Tingay} et~al.,}{{Tingay}
  et~al.}{2013}]{2013PASATingay}
{Tingay} S.~J.,  et~al., 2013, \mndoi [\pasa] {10.1017/pasa.2012.007}, \href
  {http://adsabs.harvard.edu/abs/2013PASA...30....7T} {30, e007}

\bibitem[\protect\citeauthoryear{{Tremblay} et~al.,}{{Tremblay}
  et~al.}{2015}]{2015PASATremblay}
{Tremblay} S.~E.,  et~al., 2015, \mndoi [\pasa] {10.1017/pasa.2015.6}, \href
  {http://adsabs.harvard.edu/abs/2015PASA...32....5T} {32, e005}

\bibitem[\protect\citeauthoryear{{Wang}, {Manchester}  \& {Johnston}}{{Wang}
  et~al.}{2007}]{2007MNRASWang}
{Wang} N.,  {Manchester} R.~N.,   {Johnston} S.,  2007, \mndoi [\mnras]
  {10.1111/j.1365-2966.2007.11703.x}, \href
  {http://adsabs.harvard.edu/abs/2007MNRAS.377.1383W} {377, 1383}

\bibitem[\protect\citeauthoryear{{Wayth} et~al.,}{{Wayth}
  et~al.}{2018}]{2018PASAWayth}
{Wayth} R.~B.,  et~al., 2018, \mndoi [\pasa] {10.1017/pasa.2018.37}, \href
  {https://ui.adsabs.harvard.edu/#abs/2018PASA...35...33W} {35, 33}

\bibitem[\protect\citeauthoryear{{Weisberg} et~al.,}{{Weisberg}
  et~al.}{1999}]{1999ApJSWeisberg}
{Weisberg} J.~M.,  et~al., 1999, \mndoi [The Astrophysical Journal Supplement
  Series] {10.1086/313189}, \href
  {https://ui.adsabs.harvard.edu/\#abs/1999ApJS..121..171W} {121, 171}

\bibitem[\protect\citeauthoryear{{Xue}, {Ord}, {Tremblay}, {Bhat}, {Sobey},
  {Meyers}, {McSweeney}  \& {Swainston}}{{Xue} et~al.}{2019}]{2019arXivXue}
{Xue} M.,  {Ord} S.~M.,  {Tremblay} S.~E.,  {Bhat} N.~D.~R.,  {Sobey} C.,
  {Meyers} B.~W.,  {McSweeney} S.~J.,   {Swainston} N.~A.,  2019, arXiv
  e-prints, \href {https://ui.adsabs.harvard.edu/abs/2019arXiv190500598X} {p.
  arXiv:1905.00598}

\bibitem[\protect\citeauthoryear{{Yao}, {Manchester}  \& {Wang}}{{Yao}
  et~al.}{2017}]{2017ApJYao}
{Yao} J.~M.,  {Manchester} R.~N.,   {Wang} N.,  2017, \mndoi [\apj]
  {10.3847/1538-4357/835/1/29}, \href
  {http://adsabs.harvard.edu/abs/2017ApJ...835...29Y} {835, 29}

\bibitem[\protect\citeauthoryear{{de Oliveira-Costa}, {Tegmark}, {Gaensler},
  {Jonas}, {Landecker}  \& {Reich}}{{de Oliveira-Costa}
  et~al.}{2008}]{2008MNRAS_GSM}
{de Oliveira-Costa} A.,  {Tegmark} M.,  {Gaensler} B.~M.,  {Jonas} J.,
  {Landecker} T.~L.,   {Reich} P.,  2008, \mndoi [\mnras]
  {10.1111/j.1365-2966.2008.13376.x}, \href
  {http://adsabs.harvard.edu/abs/2008MNRAS.388..247D} {388, 247}

\bibitem[\protect\citeauthoryear{{van Haarlem} et~al.,}{{van Haarlem}
  et~al.}{2013}]{2013A&AvanHaarlem}
{van Haarlem} M.~P.,  et~al., 2013, \mndoi [\aap]
  {10.1051/0004-6361/201220873}, \href
  {https://ui.adsabs.harvard.edu/#abs/2013A&A...556A...2V} {556, A2}

\bibitem[\protect\citeauthoryear{{van Straten} \& {Bailes}}{{van Straten} \&
  {Bailes}}{2011}]{2011PASAvanStraten}
{van Straten} W.,  {Bailes} M.,  2011, \mndoi [\pasa] {10.1071/AS10021}, \href
  {http://adsabs.harvard.edu/abs/2011PASA...28....1V} {28, 1}

\bibitem[\protect\citeauthoryear{{van Straten}, {Manchester}, {Johnston}  \&
  {Reynolds}}{{van Straten} et~al.}{2010}]{2010PASAvanStraten}
{van Straten} W.,  {Manchester} R.~N.,  {Johnston} S.,   {Reynolds} J.~E.,
  2010, \mndoi [\pasa] {10.1071/AS09084}, \href
  {http://adsabs.harvard.edu/abs/2010PASA...27..104V} {27, 104}

\bibitem[\protect\citeauthoryear{{van Straten}, {Demorest}  \& {Oslowski}}{{van
  Straten} et~al.}{2012}]{2012AR&TvanStraten}
{van Straten} W.,  {Demorest} P.,   {Oslowski} S.,  2012, Astronomical Research
  and Technology, \href
  {https://ui.adsabs.harvard.edu/#abs/2012AR&T....9..237V} {9, 237}

\bibitem[\protect\citeauthoryear{van~der Walt, Colbert  \& Varoquaux}{van~der
  Walt et~al.}{2011}]{2011CSEvanderWalt_numpy}
van~der Walt S.,  Colbert S.~C.,   Varoquaux G.,  2011, \mndoi [Computing in
  Science {\&} Engineering] {10.1109/MCSE.2011.37}, 13, 22

\makeatother
\end{thebibliography}

\end{document}